\documentclass[sigplan,10pt,nonacm]{acmart}
\renewcommand\footnotetextcopyrightpermission[1]{}

\usepackage[ruled,vlined]{algorithm2e}
\usepackage{algpseudocode}
\usepackage{pseudocode}
\usepackage{url}
\usepackage{amsmath}
\usepackage{float}
\usepackage{cleveref}
\usepackage{graphicx}
\usepackage{multicol}
\usepackage{multirow}
\usepackage{balance}  % for  \balance command ON LAST PAGE  (only there!)
\usepackage{tikz}
\usepackage{subcaption}
\usepackage{pgfplots}
\usepackage{pgfplotstable}
\usepackage{verbatim}
\usepackage{xspace}
\usetikzlibrary{shapes}
\usetikzlibrary{plotmarks}
\usepgfplotslibrary{units}
\usetikzlibrary{matrix}
\usepgfplotslibrary{groupplots}
\pgfplotsset{compat = newest}

\newcommand{\dbpoint}[1]{\ensuremath{{\sf x}_{#1}}}
\newcommand{\nout}[1]{N_{\rm out}({#1})}
\newcommand{\nin}[1]{N_{\rm in}({#1})}
\newcommand{\greedysearch}[4]{\ensuremath{{\rm GreedySearch}({#1},{#2},{#3},{#4})}}
\newcommand{\prune}[4]{\ensuremath{{\rm RobustPrune}({#1},{#2},{#3},{#4})}}
\newcommand{\insertpoint}[5]{\ensuremath{{\rm Insert}({#1},{#2},{#3},{#4},{#5})}}
\newcommand{\deletepoint}[3]{\ensuremath{{\rm Delete}({#1},{#2},{#3})}}

\newcommand{\insertpt}[1]{\ensuremath{{\rm Insert}({#1})}}
\newcommand{\deletept}[1]{\ensuremath{{\rm Delete}({#1})}}
\newcommand{\searchpt}[3]{\ensuremath{{\rm Search}({#1},{#2},{#3})}}

\newcommand{\cL}{\mathcal{L}\xspace}
\newcommand{\cV}{\mathcal{V}\xspace}
\newcommand{\dist}[2]{d({#1},{#2})}

\newcommand{\lti}{{\ensuremath{\sf LTI}}\xspace}
\newcommand{\deletelist}{{\ensuremath{\sf DeleteList}}\xspace}
\newcommand{\tempindex}{{\ensuremath{\sf TempIndex}}\xspace}
\newcommand{\algo}{{\ensuremath{\rm Vamana}}\xspace}
\newcommand{\merger}{{\texttt{StreamingMerge}}\xspace}
\newcommand{\algofresh}{{\textsf{FreshVamana}}\xspace}
\newcommand{\diskann}{{\ensuremath{\rm DiskANN}}\xspace}
\newcommand{\diskanntwo}{{\textsf{FreshDiskANN}}\xspace}
\newcommand{\recall}[2]{${#1}$-recall$@{#2}$}

\begin{document}
\title{FreshDiskANN: A Fast and Accurate Graph-Based ANN Index for Streaming Similarity Search}

%%
%% The "author" command and its associated commands are used to define the authors and their affiliations.

 \author{Aditi Singh}
 \email{t-adisin@microsoft.com}
 \affiliation{%
  \institution{Microsoft Research India}
 }

 \author{Suhas Jayaram Subramanya}
 \authornote{Work done while at Microsoft.}
 \email{suhasj@cs.cmu.edu}
 \affiliation{%
  \institution{Carnegie Mellon University}
 }

 \author{Ravishankar Krishnaswamy}
 \authornote{Authors listed in alphabetical order.}
 \orcid{1234-5678-9012}
 \author{Harsha Vardhan Simhadri}
 \authornotemark[2]
 \email{{rakri, harshasi}@microsoft.com}
 \affiliation{%
   \institution{Microsoft Research India}
 }

%% \author{Ben Trovato}
%% \affiliation{%
%%   \institution{Institute for Clarity in Documentation}
%%   \streetaddress{P.O. Box 1212}
%%   \city{Dublin}
%%   \state{Ireland}
%%   \postcode{43017-6221}
%% }
%% \email{trovato@corporation.com}

%% \author{Lars Th{\o}rv{\"a}ld}
%% \orcid{0000-0002-1825-0097}
%% \affiliation{%
%%   \institution{The Th{\o}rv{\"a}ld Group}
%%   \streetaddress{1 Th{\o}rv{\"a}ld Circle}
%%   \city{Hekla}
%%   \country{Iceland}
%% }
%% \email{larst@affiliation.org}

%% \author{Valerie B\'eranger}
%% \orcid{0000-0001-5109-3700}
%% \affiliation{%
%%   \institution{Inria Paris-Rocquencourt}
%%   \city{Rocquencourt}
%%   \country{France}
%% }
%% \email{vb@rocquencourt.com}

%% \author{J\"org von \"Arbach}
%% \affiliation{%
%%   \institution{University of T\"ubingen}
%%   \city{T\"ubingen}
%%   \country{Germany}
%% }
%% \email{jaerbach@uni-tuebingen.edu}
%% \email{myprivate@email.com}
%% \email{second@affiliation.mail}

%% \author{Wang Xiu Ying}
%% \author{Zhe Zuo}
%% \affiliation{%
%%   \institution{East China Normal University}
%%   \city{Shanghai}
%%   \country{China}
%% }
%% \email{firstname.lastname@ecnu.edu.cn}

%% \author{Donald Fauntleroy Duck}
%% \affiliation{%
%%   \institution{Scientific Writing Academy}
%%   \city{Duckburg}
%%   \country{Calisota}
%% }
%% \affiliation{%
%%   \institution{Donald's Second Affiliation}
%%   \city{City}
%%   \country{country}
%% }
%% \email{donald@swa.edu}

\begin{abstract} 

Approximate nearest neighbor search (ANNS) is a fundamental building
block in information retrieval with \emph{graph-based indices} being 
the current state-of-the-art~\cite{Benchmark} and widely
used in the industry. Recent advances~\cite{DiskANN19} in graph-based indices
have made it possible to index and search billion-point datasets with high
recall and millisecond-level latency on a single commodity machine with an SSD.

However, existing graph algorithms for ANNS support only \emph{static} indices 
that cannot reflect real-time changes to the corpus required by many key real-world scenarios
(e.g. index of sentences in documents, email or a news index). To overcome this drawback, 
the current industry practice for manifesting updates into such indices
is to \emph{periodically re-build these indices}, which can be
prohibitively expensive.

In this paper, we present the first graph-based ANNS index that
reflects corpus updates into the index in real-time \emph{without compromising on search performance}. 
Using update rules for this index, we design \diskanntwo, a system that can index over a
billion points on a workstation with an SSD and limited memory, and
support thousands of concurrent real-time inserts, deletes and
searches per second each, while retaining $>95\%$ \recall{5}{5}. 
This represents a 5-10x reduction in the cost of
maintaining \emph{freshness} in indices when compared to existing methods.
\end{abstract}

\settopmatter{printfolios=true,printacmref=false}
\maketitle
\pagestyle{plain}

%\maketitle

%\pagestyle{\vldbpagestyle}
%\begingroup\small\noindent\raggedright\textbf{PVLDB Reference Format:}\\
%\vldbauthors. \vldbtitle. PVLDB, \vldbvolume(\vldbissue): \vldbpages, \vldbyear.\\
%\href{https://doi.org/\vldbdoi}{doi:\vldbdoi}
%\endgroup
%\begingroup
%\renewcommand\thefootnote{}\footnote{\noindent
%This work is licensed under the Creative Commons BY-NC-ND 4.0 International License. Visit \url{https://creativecommons.org/licenses/by-nc-nd/4.0/} to view a copy of this license. For any use beyond those covered by this license, obtain permission by emailing \href{mailto:info@vldb.org}{info@vldb.org}. Copyright is held by the owner/author(s). Publication rights licensed to the VLDB Endowment. \\
%\raggedright Proceedings of the VLDB Endowment, Vol. \vldbvolume, No. \vldbissue\ %
%ISSN 2150-8097. \\
%\href{https://doi.org/\vldbdoi}{doi:\vldbdoi} \\
%}\addtocounter{footnote}{-1}\endgroup

%\ifdefempty{\vldbavailabilityurl}{}{
%\vspace{.3cm}
%\begingroup\small\noindent\raggedright\textbf{PVLDB Artifact Availability:}\\
%The source code has temporarily been made available at
%\url{https://github.com/freshdiskann/freshdiskann} and will later
%be released at an appropriate location.
%\endgroup

%%\begin{acks}
%%??
%%\end{acks}

\vspace{-10pt}
\section{Introduction} \label{sec:intro}
In the Nearest Neighbor Search problem, we are given a dataset $P$ of
points along with a pairwise distance function. The goal is to design
a data structure that, given a target $k$ and a query point $q$,
efficiently retrieves the $k$ closest neighbors for $q$ in the dataset
$P$ according to the given distance function.  This fundamental
problem is well studied in the research
community~\cite{CoverTree,babenko2014additive,Faiss17,Weber98,ECCV18,HNSW16,PQ11,Arya93,Indyk98}
and is a critical component for diverse applications in computer
vision~\cite{Wang12}, data mining~\cite{isax2}, information
retrieval~\cite{irbook}, classification~\cite{Fix89}, and
recommendation systems~\cite{DeepXML21}, to name a few.  As advances
in deep learning have made embedding-based approaches the
state-of-the-art in these applications, there has been renewed
interest in the problem at scale. Several open-source inverted-index
based search engines now support NNS~\cite{Lucene-vector,
  elasticsearch-vector, elasticsearch-vector2}, and new 
search engines based on NNS are being developed~\cite{Vespa, Milvus}.  In
newer applications of this problem, the dataset to be indexed and the
queries are the output of a deep learning model -- objects such as
sentences or images are mapped so that semantically similar objects
are mapped to closer points \cite{deep1b-link,BERT}. These points
reside in a space of dimension $d$ (typically 100-1000), and the
distance function is the Euclidean distance ($\ell_2$) or cosine
similarity (which is identical to $\ell_2$ when the data is
normalized).

Since it is impossible to retrieve the exact nearest neighbors without
a cost linear in the size of the dataset in the general case 
(see~\cite{Indyk98,Weber98}) due to a phenomenon known as the
\emph{curse of dimensionality}~\cite{Clarkson94}, one aims to find the
\emph{approximate nearest neighbors} (ANN) where the goal is to
retrieve $k$ neighbors that are close to being optimal.  The quality
of an ANN algorithm is judged by the \emph{trade-off} it provides
between accuracy and the hardware resources such as compute, memory
and I/O consumed for the search.

Even though this abstraction of ANN search is widely studied, it does not capture many important
real-world scenarios where user interactions with a system creates and destroys data, and results in updates to
$P$ (especially in the literature on graph-based ANNS
indices~\cite{GraphANNSSurvey21}). % @ravi: What does this mean?
For example, consider an enterprise-search scenario where the system 
indexes sentences in documents generated by users across
an enterprise. Changes to sentences in a document would
correspond to a set of new points inserted and previous points
deleted. Another scenario is an email server where arrival and deletion of emails 
correspond to insertion and deletion of points into an ANNS index. ANNS systems 
for such applications would need to host indices
containing trillions of points with real-time updates that can 
reflect changes to the corpus in user searches, ideally in real-time.

Motivated by such scenarios, we are interested in solving the {\bf
  fresh-ANNS} problem, where the goal is to support ANNS on a 
  \emph{continually changing set of points}.  
  %This is in contrast to 
  %the classical \emph{static} ANNS problem that requires the set $P$ 
  %to be \emph{fixed}. 
  Formally, we define the fresh-ANNS
problem thus: given a time varying dataset $P$ (with state $P_t$ at
time $t$), the goal is to maintain a dynamic index that computes the
approximate nearest neighbors for any query $q$ issued at time $t$ only on
the active dataset $P_t$. Such a system must support three operations
(a) \emph{insert} a new point, (b) \emph{delete} an existing point, and (c) \emph{search}
for the nearest neighbors given a query point. The overall quality of a fresh-ANNS system is
measured by:
\begin{itemize}
	\item The recall-latency tradeoff for search queries, and its
	robustness over time as the dataset $P$ evolves.
	\item Throughput and latency of  insertions and deletions.
	\item Overall hardware cost (CPU, RAM and SSD footprint) to build and maintain such an index.
\end{itemize}

We are interested in \emph{quiescent
  consistency}~\cite{HerlihyShavit12, QuiscentDef14}, where the
results of search operations executed at any time $t$ are consistent
with some total ordering of all insert and delete operations completed
before $t$.

We use the following notion of recall in this paper.\footnote{ An
  index that provides good \recall{k}{k} can be used to satisfy other
  notions of recall such as finding all neighbors within a certain
  radius.}
\vspace{-3pt}
\begin{definition}[{\bf \recall{k}{k}}]
  \label{def:recall}
  For a query vector $q$ over dataset $P$, suppose that (a) $G
  \subseteq P$ is the set of actual $k$ nearest neighbors in $P$, and
  (b) $X \subseteq P$ is the output of a $k$-ANNS query to an index. Then the \recall{k}{k} for the
  index for query $q$ is $\frac{|X \cap G|}{k}$. Recall for a set of
  queries refers to the average recall over all queries.
\end{definition}
\vspace{-3pt}

\textbf{Goal.} Motivated by real-world scenarios, we seek to build the
most cost-effective system for the fresh-ANNS problem which can
maintain a billion-point index using commodity machines with 128GB RAM
and a 2TB SSD\footnote{Henceforth, when we refer to ``a machine'', we
  implicitly refer to this configuration unless otherwise specified. }
and support \emph{thousands} of real-time inserts and deletes per
second, and also \emph{thousands} of searches per second with high
accuracy of 95+\% \recall{5}{5}.  Indeed, the current state-of-art
system for fresh-ANNS which can support comparable update and search
performance on a billion-point dataset is based on the classical LSH
algorithm~\cite{Sundaram13}, and requires a hundred machines of 32GB
RAM (translating to around 25 machines of our stated
configuration). \emph{In this work, we seek to reduce this deployment
  cost down to a single machine per billion points.} To handle
trillion-point indices (as in web-search scenarios), one can employ a
simple distributed approach wherein thousand machines host a billion
points each -- queries are broadcast and results aggregates while
updates are routed to the appropriate nodes.

\vspace{-5pt}
\subsection{Shortcoming of existing algorithms}

Of all the algorithms for static-ANNS, the ones most easily capable of
supporting streaming support are the ones based on simple hashing
algorithms such as LSH (locality sensitive hashing). However, these
algorithms suffer from either being too memory intensive, needing to
store hundreds of hash functions in main memory, or become extremely
slow for query processing when the index is stored on secondary
storage. For example, the state-of-art system for streaming similarity
search (or fresh-ANNS), PLSH~\cite{Sundaram13}, is a parallel and
distributed LSH-based mechanism. While it offers comparable update
throughput and search performance as our system, it ends up needing
25X more machines due to the high RAM consumption. A similar issue can
be seen with PM-LSH, another state-of-art system based on
LSH~\cite{Zheng20}, where the memory footprint is a bit lower than
PLSH (due to the system using fewer LSH tables), but the query
latencies are an order of magnitude slower than our system and
PLSH. Alternately, disk-based LSH indices such as SRS~\cite{Sun14} can
host a billion-point index on a single machine, but the query
latencies are extremely slow with the system fetching around 15\% of
the total index (running into GBs per query) from the disk to provide
good accuracy. Another recent algorithm HD-Index~\cite{Arora18} can
serve a billion-point index with just a few megabytes of RAM
footprint, but it suffers from search latencies of a few seconds to
get accuracy of around $30\%$. Moreover, the algorithm only handles
insertions, and simply performs a variant of blacklisting for
deletions, and hence would need periodic rebuilding. Finally, there
are other classes of ANNS algorithms such as kd-Tree~\cite{Bentley75},
Cover Trees~\cite{Beygelzimer06} which support reasonably efficient
update policies, but these algorithms work well only when the data
dimensionality is moderately small (under 20); their performance drops
when the data dimensionality is 100 or more which is typical for
points generated by deep-learning models..

At the other end of the spectrum of ANNS indices are \emph{graph-based
  indexing algorithms}~\cite{HNSW16,NSG17,NGT1,NGT2,NGT3,DiskANN19}.
Several comparative
studies~\cite{Benchmark,Echihabi19,Li20,GraphANNSSurvey21} of ANNS
algorithms have concluded that they significantly out-perform other
techniques in terms of search throughput on a range of real-world
static datasets. These algorithms are also widely used in the industry
at scale. However, \emph{all known graph indices are static and do not
  support updates, especially delete requests}~\cite{HNSW-rebuild},
possibly due to the fact that simple graph modification rules for
insertions and deletions do not retain the same graph quality over a
stream of insertions and deletions.

As a result, the current practice in industry is to periodically
re-build such indices from scratch~\cite{HNSW-rebuild} to manifest
recent changes to the underlying dataset. However, this is a very
expensive operation. It would take about 1.5-2 hours on a dedicated
high-end 48-core machine to build a good quality HNSW
index~\cite{HNSW-git} over 100M points. So we would need \emph{three
  dedicated machines} for constantly rebuilding indices to maintain
even \emph{six-hourly freshness guarantee over a billion-point
  index}. This is apart from the cost of actually serving the indices,
which would again be anywhere between one for DRAM-SSD hybrid
indices~\cite{DiskANN19} to four for in-memory indices~\cite{HNSW-git}
depending on the exact algorithm being deployed. This paper aims to
serve and update an index over a billion points with \textbf{real-time
  freshness} using just one machine. This represents a significant
cost advantage for web and enterprise-scale search platforms that need
to serve indices spanning trillions of points.

\vspace{-10pt}
\subsection{Our Contributions}
In this paper, we present the \textbf{\diskanntwo} system to solve the
fresh-ANNS problem for points in Euclidean space with real-time
freshness, and with 5-10x fewer machines than the current
state-of-the-art. As part of this, we make several technical
contributions:
\begin{enumerate}
	\item We demonstrate how simple graph update rules result in degradation of index quality over a stream of insertions and deletions for popular graph-based algorithms such as HNSW~\cite{HNSW16} and NSG~\cite{NSG17}. 
	\item We develop \algofresh, the first 
		graph-based index that supports insertions and deletions, and empirically demonstrate its stability over long streams of updates.
	\item In order to enable scale, our system stores the bulk of the graph-index on an SSD, with only the most recent updates stored in memory. To support this, we design a novel two-pass \merger algorithm which makes merges the in-memory index with the SSD-index in a very write-efficient manner (crucial since burdening the SSD would lead to worse search performance as well). Notably, the \emph{time and space complexity of the merge procedure is proportional to the change set}, thereby making it possible to update large billion-point indices on a machine with limited RAM using an order of magnitude less compute and memory than re-building the large index from scratch. 
	\item Using these ideas, we design the \diskanntwo system to
	consist of a long-term \emph{SSD-resident index} over the majority of the
	points, and a short-term \emph{in-memory index} to aggregate recent
	updates. Periodically, unbeknownst to the
	end user, \diskanntwo consolidates the short-term index into the long-term index using our \merger process in the background to bound the memory footprint of the short-term index, and hence the overall system.
\end{enumerate}

We conduct rigorous week-long experiments of this system on an (almost) billion
point subset of the popular SIFT1B~\cite{SIFT1B} dataset on a 48 core machine and 3.2TB SSD. We monitor recall
stability, end-user latency and throughput for updates and
searches. Some highlights are:

\begin{itemize}
	
	\item The system uses less than 128GB of DRAM at all times.
	\item The \merger can merge a 10\% change to the index
	(5\% inserts + 5\% deletes) to a billion-scale index in
	$\sim$10\% of the time than it takes to rebuild the index.
	\item \diskanntwo can support a steady-state throughput of 1800 inserts and 1800 deletes per
	second while retaining freshness and without backlogging
	background merge. The
	system can also support short bursts of much higher change rate, up to even 40,000 inserts/second. 
	\item The user latency of insertion and deletion is under 1ms, even when
	a background merge is underway. 	
	\item \diskanntwo supports 1000 searches/sec with 95+\% \recall{5}{5}
	over the latest content of the index, with mean search latency well under $20ms$.	
\end{itemize}

\section{Related Work}
\label{sec:recap}

ANNS is a classical problem with a large body of research work.
Recent surveys and benchmarks~\cite{Benchmark,Echihabi19,Li20} provide
a great overview and comparison of the state-of-the-art ANN
algorithms. This section focuses on the algorithms relevant for
vectors in high-dimensional space with Euclidean metrics, and examines
their suitability for the fresh-ANNS setting we consider in this
paper.  Beyond ANNS for points in Euclidean spaces, there has been
work for tailored inputs and other notions of similarity such as those
for time series data, e.g., ~\cite{isax2,coconut,Agrawal93}. The
work~\cite{Echihabi19} provides a comprehensive study of such
algorithms and their applicability. %In this paper, we limit our study
%to Euclidean spaces of 100-1000 dimensions which are commonly
%used for deep-learning based search systems.

{\bf Trees.} Some of the early research on ANNS focused on low-dimensional points
(say, $d \leq 20$). For such points, spatial
partitioning ideas such as $R^*$-trees~\cite{Beckmann90},
kd-trees~\cite{Bentley75} and Cover Trees~\cite{CoverTree} work well,
but these typically do not scale well for high-dimensional data owing
to the curse of dimensionality. There have been some recent advances
in maintaining several trees and combining them with new ideas to
develop good algorithms such as FLANN~\cite{FLANN} and
Annoy~\cite{github:annoy}. However, they are built for static indices,
and moreover, even here, the graph-based algorithms outperform
them~\cite{Benchmark} on most datasets.

{\bf Hashing.} In a breakthrough result, Indyk and
Motwani~\cite{Indyk98} show that a class of algorithms, known as
\emph{locality sensitive hashing} can yield provably approximate
solutions to the ANNS problem with a polynomially-sized index and
sub-linear query time. Subsequent to this work, there has been a
plethora of different LSH-based algorithms~\cite{LSHSurvey08,Indyk98,Zheng20}, including
those which depend on the data~\cite{LSHDataDep15}, use spectral
methods~\cite{Weiss08}, distributed LSH~\cite{Sundaram13}, etc. While
the advantage of the simpler data-independent hashing methods are that
updates are almost trivial, the indices are often entirely resident in
DRAM and hence do not scale very well. Implementations which make use of auxiliary storage such as
SRS~\cite{Sun14} typically have several orders of magnitude slower
query latencies compared to the graph-based algorithms. 
Other hashing-based methods~\cite{NSH15, ASH11, SGH15} learn an
optimal hash family by exploiting the neighborhood graph. Updates to
an index would require a full re-computation of the family and hashes
for every database point, making them impractical for fresh-ANNS.

{\bf Data quantization} and {\bf Inverted indices} based algorithms
have seen success w.r.t the goal of scaling to large datasets with low
memory footprint. These algorithms effectively reduce the
dimensionality of the ANNS problem by \emph{quantizing} vectors into a
compressed representation so that they may be stored using smaller
amount of DRAM. Some choices of quantizers~\cite{Faiss17} can support
GPU-\emph{accelerated} search on billion-scale datasets. Popular
methods like IVFADC~\cite{PQ11}, OPQ~\cite{opq-pami},
LOPQ~\cite{lopq}, FAISS~\cite{Faiss17},
IVFOADC+G+P ~\cite{BillionInvIdx19} and IMI~\cite{IMI12} exploit the
data distribution to produce low memory-footprint indices with
reasonable search performance when querying for a large number of
neighbors. While most methods\cite{babenko2014additive,Faiss17,
  PQ11,opq-pami} minimize the vector \emph{reconstruction error} $||x
- x^\dagger||^2$, where $x$ is a database vector and $x^\dagger$ is
its reconstruction from the quantized representation, Anisotropic
Vector Quantization ~\cite{guoaccelerating} optimizes for error for
maximum inner-product search. Some of these systems such as
FAISS~\cite{Faiss17} support insert and delete operations on an
existing index under reasonable conditions like stationary data
distributions. However, due to the irreversible loss due to the
compression/quantization, these methods fail to achieve even moderate
values of \recall{1}{1}, sometimes plateauing at $50\%$ recall. These
methods offer good guarantees on weaker notions such as
\recall{1}{100}, which is the likelihood that the true nearest
neighbor for a query appears in a list of $100$ candidates output by
the algorithm. Hence they are not the methods of choice for
high-recall high-throughput scenarios. 

A recent work, ADBV ~\cite{AnalyticDBV}, proposes a hybrid model for 
supporting {\bf streaming inserts and deletes}. New points are inserted into 
an in-memory HNSW~\cite{HNSW16} index while the main on-disk 
index utilises a new PQ-based indexing algorithm called VGPQ. In order to mitigate the 
accuracy loss due to PQ, VGPQ search performs a large number of 
distance computations and incurs high search latencies. 
As distributed system over several powerful nodes, 
the model has low search throughput even when no inserts and deletes 
are going on. Hence, such a system cannot be used in high-throughput scenarios.

A recent work, ADBV ~\cite{AnalyticDBV}, proposes a hybrid SQL-vector search model.
New vectors are inserted into an in-memory HNSW index while the main on-disk index
 spanning upto a billion points is spread across multiple machines.
The on-disk index is an extension of IVF-clustering~\cite{PQ11}
which is far less efficient for search compared to graph indices
in terms of the number of distance comparisons and I/O.
As a result, their aggregate search throughput on a billion point index 
spread across disks on 16 machines is lesser than the throughput of 
\diskanntwo with one machine. Our work achieves this by designing an 
on-SSD updatable graph index which is far more efficient for search.
Their insertion throughput on an index spread across 70 machines is 
also much lesser than that of \diskanntwo on one machine.

%%  {\bf Graph-based algorithms}, as of this writing, are the best
%%  performing algorithms for high-recall ANNS on real-world datasets
%%  ~\cite{Benchmark}. Algorithms like HNSW~\cite{HNSW16},
%%  NSG~\cite{NSG17}, NGT~\cite{NGT1,NGT2,NGT3} and
%%  \algo~\cite{DiskANN19} construct a \emph{navigable} graph whose
%%  vertex set is the set of base points i.e. each base point is
%%  represented by a node in the graph. To the best of our knowledge,
%%  none of these algorithms support dynamic updates to the index, as
%%  handling delete requests without altering graph quality is a
%%  non-trivial task; see~\Cref{subsec:algo_hnsw_cmp} for a brief
%%  discussion. While most of these graph-based indices are resident in
%%  main memory, \diskann~\cite{DiskANN19} leverages a combination of a
%%  graph index called \algo optimized for serving from SSD, along with a
%%  quantization scheme known as Product Quantization~\cite{PQ11} to
%%  maintain large-scale indices with low memory footprint. On
%%  billion-point datasets, \diskann can provide significantly higher
%%  \recall{k}{k} with the same memory and latency budget as FAISS.

%%  Another use case is the training of deep-learning models
%% training where ANNS indices are used to keep track of the most
%% relevant activations to consider, which changes as backpropagation
%% proceeds~\cite{}.

\section{Graph-based ANNS indices} \label{sec:graph-recap}

In this section, we recap how most state-of-the-art graph-based indices work for static-ANNS and also highlight the issues they face with supporting deletions. 

\vspace{-6pt}
\subsection{Notation}

The primary data structure in graph indices is a directed graph with
vertices corresponding to points in $P$, the dataset that is
to be indexed, and edges between them. With slight notation overload,
we denote the graph $G = (P,E)$ by letting $P$ also denote the vertex
set.  Given a node $p$ in this directed graph, we let $\nout{p}$ and
$\nin{p}$ denote the set of out- and in-edges of $p$.  We denote the
number of points by $n = |P|$.  Finally, we let $\dbpoint{p}$ denote
the database vector corresponding to $p$, and let $\dist{p}{q} = ||\dbpoint{p}
- \dbpoint{q}||$ denote the $\ell_2$ distance between two points $p$ and $q$.
We now describe how graph-based ANNS indices are built and used for
search.

\begin{algorithm}[t]
	\DontPrintSemicolon \small \KwData{ Graph $G$ with start node
          $s$, query \dbpoint{q}, result size $k$, search list size $L
          \geq k$}
	
	\KwResult{Result set $\cL$ containing $k$-approx NNs, and a set $\cV$ containing all the visited nodes}
	
	\tt
	\Begin{
		initialize sets $\cL\gets \{s\}$ and $\cV\gets\emptyset$\;
		\While{$\cL \setminus \cV \neq \emptyset$}{
			let $p* \gets \arg \min_{p \in \cL\setminus \cV} || \dbpoint{p} - \dbpoint{q}||$\;
			update $\cL \gets \cL \cup \nout{p^*}$ and $\cV \gets \cV \cup \{p^*\}$\;
			\If{$|\cL| > L$} {
				update $\cL$ to retain closest $L$ points to \dbpoint{q}\;
			}
		}
		return $[$closest $k$ points from $\cV$; $\cV$$]$\;
	}
	\caption{\greedysearch{s}{\dbpoint{q}}{k}{L}}
	\label{alg:greedysearch}
\end{algorithm}

\subsection{Navigability and Index Search}

Roughly speaking, navigability of a directed graph is the property
that ensures that the index can be queried for nearest neighbors using
a \emph{greedy} search algorithm. The greedy search algorithm
traverses the graph starting at a designated \emph{navigating} or
start node $s\in P$.  The search iterates by greedily walking
from the current node $u$ to a node $v\in \nout{u}$ that minimizes the distance
to the query, and terminates when it reaches a \emph{locally-optimal}
node, say $p^*$, that has the property $d(p^*, q) \leq
d(p, q)\,\forall p\in\nout{p^*}$. Greedy search cannot improve
distance to the query point by navigating \emph{out} of $p^*$ and
returns it as the candidate nearest neighbor for query
$q$. \Cref{alg:greedysearch} describes a variant of this greedy search
algorithm that returns $k$ nearest neighbor candidates.

\noindent\textbf{Index Build} consists of constructing a navigable
graph.  The graph is typically built to achieve two contrasting
objectives to minimize search complexity: (i) make the greedy search
algorithm applied to each base point $p \in P$ in the vertex set
converge to $p$ in the fewest iterations (intuitively, this would ensure that~\Cref{alg:greedysearch}
converges to $p$ when searching for a query $\dbpoint{q}$ if $p$ is the
nearest-neighbor for $\dbpoint{q}$), and
(ii) have a maximum out-degree of at most $R$ for all $p\in P$, a
parameter typically between $16-128$.  

Algorithms like
NN-\emph{Descent}~\cite{approx-kann-11} use gradient descent
techniques to determine $G$. Others start with a specific type of
graph --- an empty graph with no edges~\cite{HNSW16,DiskANN19} or an approximate
$k-$NN graph~\cite{NSG17, EFANNA} --- and iteratively refine $G$ using the following two-step
construction algorithm to improve navigability:
\begin{itemize}
	\item \textbf{Candidate Generation} - For each base point
          $\dbpoint{p}$, run \Cref{alg:greedysearch} on $G$ to obtain
          $\mathcal{V}, \mathcal{L}$. $\mathcal{V}\cup\mathcal{L}$
          contains nodes \emph{visited} and/or closest to $p$ in $G$
          during the search in the current graph $G$, making them good
          candidates for adding to $\nout{p}$ and $\nin{p}$, thereby
          improving the navigability to $p$ in the updated graph $G$.
        
	\item \textbf{Edge Pruning} -- When the out-degree of a node
	$p$ exceeds $R$, a pruning algorithm
	(like \Cref{alg:robustprune} with $\alpha$ set to $1$) filters out similar kinds of (or
	redundant) edges from the adjacency list to ensure $|\nout{p}|\leq R$. Intuitively, the procedure sorts the neighbors of $p$ in increasing order of distance from $p$, and only retains an edge $(p,p'')$ if there is no edge $(p,p')$ which has been retained and $p'$ is closer to $p''$ than $p$ (i.e., if \Cref{alg:greedysearch} can reach $p''$ from $p$ through $p'$, then we can safely remove the edge $(p,p'')$).
\end{itemize}

\subsection{Why are Deletions Hard?} \label{sec:degrading}
While graph-indices offer state-of-the-art search performance, all known algorithms apply for the static-ANNS problem. In particular, deletions pose a big challenge for all these algorithms -- e.g., see this discussion~\cite{HNSW-rebuild} on HNSW supporting delete requests by adding them to a blacklist and omitting from search results. Arguably, this is due to the lack of methods which modify the navigable graphs while retaining the original search quality. To further examine this phenomenon, we considered three popular static-ANNS algorithms, namely HNSW, NSG, and \algo and tried the following natural update policies when faced with insertions and deletions. 

\smallskip \noindent {\bf Insertion Policy.} For insertion of a new point $p$, we run the candidate generation algorithm as used by the respective algorithms and add the chosen in- and out-edges, and if necessary, whenever the degree of any vertex exceeds the budget, run the corresponding pruning procedure.

\smallskip \noindent {\bf Delete Policy A.} When a point $p$ is deleted, we simply remove all in- and out-edges incident to $p$, without adding any newer edges to compensate for potential loss of navigability. Indeed, note that $p$ might have been on several navigating paths to other points in the graph.

\smallskip \noindent {\bf Delete Policy B.} When a point $p$ is deleted, we remove all in- and out-edges incident to $p$, and add edges in the local neighborhood of $p$ as follows: for any pair of directed edges $(p_{\rm in}, p)$ and $(p, p_{\rm out})$ in the graph, add the edge $(p_{\rm in}, p_{\rm out})$ in the updated graph. If the degree bound of any vertex is violated, we run the pruning procedure associated with the respective algorithm to control the degrees.

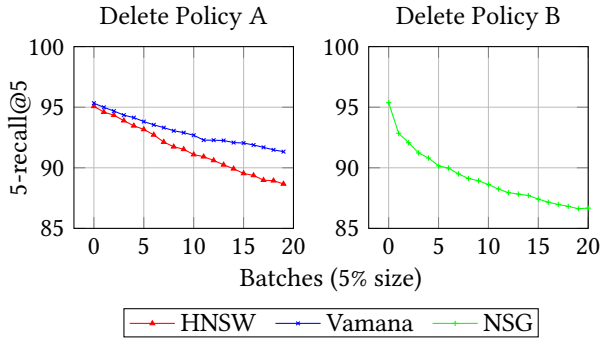
\begin{figure}[t]
  \begin{center}
  \begin{tikzpicture}
    \begin{groupplot}[group style={group size= 2 by 1}, height=4cm, width=4.5cm]
      \hspace{-10pt}
        \nextgroupplot[title=Delete Policy A, title style={yshift=-0.8ex}, ylabel={5-recall@5}, ylabel shift=-8pt,ymin=85, ymax=100, xmax=20, grid=major]
        \addplot [red, mark = triangle*, mark size = 1pt]table[x=X,y=Y,col sep=space] {plots/Hnsw_RemDelRef_sift1m_ss5_recall.txt};\label{hnsw_del_curve}
        \addplot [blue, mark = x, mark size = 1pt]table[x=X,y=Y,col sep=space] {plots/Vamana_RemDelRef_sift1m_ss5_recall.txt}; \label{vamana_del_curve}
        \coordinate (top) at (rel axis cs:0,1);% coordinate at top of the first plot
        \nextgroupplot[title=Delete Policy B, title style={yshift=-0.8ex}, ymin=85, ymax=100, xmax=20, grid=major]
        \addplot [green, mark = +, mark size = 1pt]table[x=X,y=Y1,col sep=space] {plots/Lazy_reinc_nsg_alpha_1_5.txt}; \label{nsg_del_curve}
        \coordinate (bot) at (rel axis cs:1,0);% coordinate at bottom of the last plot
    \end{groupplot}
    \path (top|-current bounding box.north) --
      coordinate(legendpos)
      (bot|-current bounding box.north);
\matrix[
    matrix of nodes,
    anchor=south,
    draw,
    inner sep=0.1em,
    column sep=0em,
    draw
  ]at([yshift=-30ex]legendpos)
  {
    \ref{hnsw_del_curve}& HNSW&[5pt]
    \ref{vamana_del_curve}& Vamana&[5pt]
    \ref{nsg_del_curve}& NSG\\};
\matrix[
    matrix of nodes,
    anchor=south,
    inner sep=0em,
  ]at([yshift=-26ex]legendpos)
  {Batches (5\% size)\\};
    \end{tikzpicture}
  \vspace{-8pt}
  \caption{Search recall over 20 cycles of deleting and re-inserting 5\% of SIFT1M dataset
  with statically built HNSW, \algo, and NSG indices with $L_s$ = 44, 20, 27, respectively.
  \vspace{-10pt}}
  \label{fig:recall-drop-deletes}
\end{center}
\end{figure}

\Cref{fig:recall-drop-deletes} shows that both of these delete policies
are not effective. In this experiment, we consider the SIFT1M
dataset~\cite{sift-link} comprising of a million points in $128$
dimensions, and start with the static-ANNS index for each of the
algorithms. We then compose an update stream by selecting 5\% of the
points at random and deleting them, followed by presenting them again
as insertions. We then repeat this process over multiple cycles. A
stable update policy should result in similar search performance after
each cycle since the index is over the same dataset. However, all of
the algorithms show a consistently deteriorating trend in search
performance (the recall drops for a fixed candidate list size). The
left plot in~\Cref{fig:recall-drop-deletes} shows the trend for HNSW
and \algo indices with Delete Policy A, while the other considers the
Delete Policy B for the NSG index. Other combinations show similar
trends but we omit them due to lack of space.

\vspace{-6pt}
\section{The  $\boldsymbol{\algofresh}$ algorithm}
\label{sec:graph-updates}

Following the experiments in~\Cref{sec:degrading}, we investigated the
 reason that the recall drops over multiple cycles of updates for
 deleting and re-inserting the same set of points. It turns out that
 the graph becomes sparse (lesser average degree) as we update it,
 and hence it becomes less navigable. We suspect that this is due to
 the very aggressive pruning policies of existing algorithms such as
 HNSW and NSG use to favor highly sparse graphs.
%While this is sufficient for static indices since the quality of the candidate set on which they run the prune procedure is of high quality, this is no longer the case when we make local modifications as in delete policy B, where the quality of candidate set is not sufficiently diverse to prevent many edges from being pruned.

Fortunately, the sparsity-vs-navigability issue has recently been
studied from a different perspective in~\cite{DiskANN19}, where the
authors seek to build denser graphs to ensure the navigating paths
converge much quicker. This in turn enables them to store such graphs
on the SSD and retrieve the neighborhood information required
by~\Cref{alg:greedysearch} as required from the SSD without incurring
large SSD latencies.

\medskip \noindent {\bf $\boldsymbol{\alpha}$-RNG Property.} The crucial idea in the graphs constructed in~\cite{DiskANN19} is a more relaxed pruning procedure, which removes an edge $(p, p'')$ only if there is an edge $(p,p')$ and $p'$ must be significantly closer to $p''$ than $p$, i.e., $d(p', p'') < \frac{d(p, p'')}{\alpha}$ for some $\alpha > 1$. Generating such a graph using $\alpha > 1$ intuitively ensures
that the distance to the query vector progressively decreases
geometrically in $\alpha$ in~\Cref{alg:greedysearch} since we remove
edges only if there is a detour edge which makes significant progress
towards the destination. Consequently, the graphs become denser as $\alpha$
increases.

\emph{We now present one of our crucial findings and contributions --  
   graph index update rules for insertions and deletions that exploit
  the $\alpha$-RNG property to ensure continued navigability of the
  graph and retain stable recall over multiple modifications.}

%At a high level, we implement the natural algorithm for insertions (candidate generation followed by pruning), and for deletions, we add all edges in the local neighborhood of the point as in Delete Policy B from~\Cref{sec:degrading}, and follow it with the pruning procedure. However, we crucially use $\alpha> 1$ for the pruning process for insertions as well as deletions. 

%Our new algorithm \algofresh builds on the \algo index, and allows
%efficient insertions and deletions of points.  For
%clarity, we describe the update rules as applied to a graph index
%stored entirely in memory, and defer discussion on potential SSD-based
%deployments to~\Cref{sec:final-design}.  Further, in
%~\Cref{subsec:algo_hnsw_cmp}, we show that applying analogous update
%rules to prior graph indices (without the $\alpha$-RNG property)
%results in graphs with progressively deteriorating quality.

 %% We then present a discussion on how analogous update rules would fare
 %% on other state-of-the-art graph indices which do not exploit the
 %% $\alpha$-RNG property in~\Cref{subsec:hnsw_cmp}. We validate these
 %% update rules by presenting rigorous experiments showing the recall
 %% stability in~\Cref{subsec:algofresh_recall_stability}.

\begin{algorithm}[t]
  \DontPrintSemicolon
  \small
  \KwData{Graph $G(P,E)$ with start node $s$, new point to be added with vector $\dbpoint{p}$, distance threshold $\alpha > 1$, out degree bound $R$, search list size $L$}
  \KwResult{Graph $G'(P',E')$ where $P' =  P \cup \{p\}$}
  \tt
  \Begin{
      initialize set of expanded nodes $\cV \gets \emptyset$\;
      initialize candidate list $\mathcal{L} \gets \emptyset$\;
      $[\mathcal{L},\cV] \gets \greedysearch{s}{p}{1}{L}$        \;
      set $p$'s out-neighbors to be $\nout{p} \gets \prune{p}{\cV}{\alpha}{R}$ (\Cref{alg:robustprune}) \;
      \ForEach{$j \in \nout{p}$}{
        \eIf{$|\nout{j} \cup \{p\}| > R$}
            {
              $\nout{j} \gets \prune{j}{\nout{j} \cup \{p\}}{\alpha}{R}$\;
            }{
              update $\nout{j} \gets \nout{j} \cup \{p\}$\;
            }
      }
    }
    \caption{\insertpoint{\dbpoint{p}}{s}{L}{\alpha}{R}}
    \label{alg:insert}
\end{algorithm}

\begin{algorithm}[t]
	\DontPrintSemicolon \small \KwData{ Graph $G$, point $p \in
		P$, candidate set $\cV$, distance threshold $\alpha\geq 1$,
		degree bound $R$ } \KwResult{$G$ is modified by setting at
		most $R$ new out-neighbors for $p$}
	
	\tt
	\Begin{
		$\cV \gets (\cV \cup \nout{p}) \setminus \{p\}$\;
		$\nout{p} \gets \emptyset$\;
		\While{$\cV \neq \emptyset$}{
			$p^* \gets \arg \min_{p' \in \cV} d(p,p')$\;
			$\nout{p} \gets \nout{p} \cup \{p^*\}$\;
			\If{$|\nout{p}| = R$} {
				break\;
			}
			\For{$p' \in \cV$}{
				\If{$\alpha \cdot d(p^*, p') \leq d(p,p')$} {
					remove $p'$ from $\cV$\;    
				}
			}	
			
		}
	}
	\caption{\prune{p}{\cV}{\alpha}{R}}
	\label{alg:robustprune}
\end{algorithm}

\vspace{-6pt}
\subsection{Insertion} \label{sec:insert}
A new point $\dbpoint{p}$ is inserted into a \algofresh index
using~\Cref{alg:insert}. Intuitively, it queries the current index for
nearest neighbors of $p$ to obtain the visited set $\cV$, generates
candidate out-neighbors for $\dbpoint{p}$ using pruning procedure
in~\Cref{alg:robustprune} on $\cV$, and adds bi-directed edges between
$p$ and the pruned candidates. If out-degree of any vertex exceeds
$R$, \Cref{alg:robustprune} can be used to prune it to $R$.

We use lock-based concurrency control to guard access to $\nout{p}$
for a node $p$, allowing for high insertion throughput using multiple
threads. Due to the fine granularity of locking and the short duration
for which the locks are held, insertion throughput scales
near-linearly with threads (see Appendix).

%% \footnote{Indeed, even
%%   the way the static graph was constructed was largely incremental,
%%   with the algorithm iterating over points in $P$ and refining the
%%   graph over time; see~\Cref{sec:recap} for a high-level
%%   explanation.}

\begin{algorithm}[t]
  \DontPrintSemicolon
  \small
\KwData{Graph $G(P,E)$ with $|P| = n$, set of points to be deleted $L_D$}
\KwResult{Graph on nodes $P'$ where $P' = P \setminus L_D$}

\tt
\Begin
{
  \ForEach{$p \in P \setminus L_D$ s.t. $\nout{p} \cap L_D \neq \emptyset$}{
    $\mathcal{D} \gets \nout{p} \cap L_D$\;
             $\mathcal{C} \gets \nout{p} \setminus \mathcal{D}$  //initialize candidate list \;
            \ForEach{$v \in \mathcal{D}$}{
            {
            $\mathcal{C}\gets \mathcal{C}\cup \nout{v}$\;
            }
            }
            $\mathcal{C} \gets \mathcal{C} \setminus \mathcal{D} $\;
            $\nout{p} \gets \prune{p}{\mathcal{C}}{\alpha}{R}$\;
        }
}
\caption{\deletepoint{L_D}{R}{\alpha}}
\label{alg:deletion}
\end{algorithm}

\vspace{-6pt}
\subsection{Deletion} \label{sec:delete}
%A naive way of deleting a point \dbpoint{p} would be to remove the
%corresponding node $p$ from the graph. However, when a large number of
%deletes are processed in this manner, the graph becomes less navigable
%due to missing edges, and many nodes become unreachable from the start
%node $s$ using~\Cref{alg:greedysearch}, thereby degrading the index quality. 
Our deletion algorithm~\Cref{alg:deletion} is along the lines of Delete Policy B in~\Cref{sec:degrading}, with the crucial feature being using the relaxed $\alpha$-pruning algorithm to retain density of the modified graph. 
%We avoid this by \emph{adding suitable edges} in the local
%neighborhoods of a deleted node to preserve navigability.
Specifically, if $p$ is deleted, we add edges $(p', p'')$ whenever $(p',p)$ and $(p,p'')$ are directed edges in the current graph.  
%Since all search paths through $p$ must follow such a path $p' \rightarrow p \rightarrow p''$, add such edges $(p',p'')$ should preserve all search paths. 
 In this
process, if $|\nout{p'}|$ exceeds the maximum out-degree $R$, we prune
it using~\Cref{alg:robustprune}, preserving the $\alpha-$RNG property.

However, since this operation involves editing the neighborhood for
all the in-neighbors of $p$, it could result be expensive to
do \emph{eagerly}, i.e., processing deletes as they arrive. \algofresh employs a \emph{lazy} deletion strategy -- when a point $p$
is deleted, we add $p$ to a \deletelist without changing the
graph. \deletelist contains all the points that have been deleted
but are still present in the graph. At search time, a modified~\Cref{alg:greedysearch} 
uses nodes in the \deletelist for navigation, but filters them out from the result set.

\textbf{Delete Consolidation.} After accumulating a non-trivial number
of deletions (say 1-10\% of the the index size), we batch-update the
graph using~\Cref{alg:deletion} to update the neighborhoods of points
with out-edges to these deleted nodes.  This operation is trivially
parallelized using prefix sums to consolidate the vertex list, and a
parallel map operation to locally update the graph around the deleted
nodes.

%Using the \emph{local} update rules discussed above, for each deleted node $p$ in the \deletelist,
%we remove references to $p$, append $\nout{p}$ to $\nout{p'}$, and
%prune the resulting list to size $R$ for all $p'$ with out-edges to $p$. 

\subsection{Recall stability of \algofresh} \label{subsec:algofresh_recall_stability}
We now demonstrate how using our insert and delete algorithms (along with a choice of $\alpha > 1$) ensures that the resulting index is stable over a long stream of updates. %To this end, 

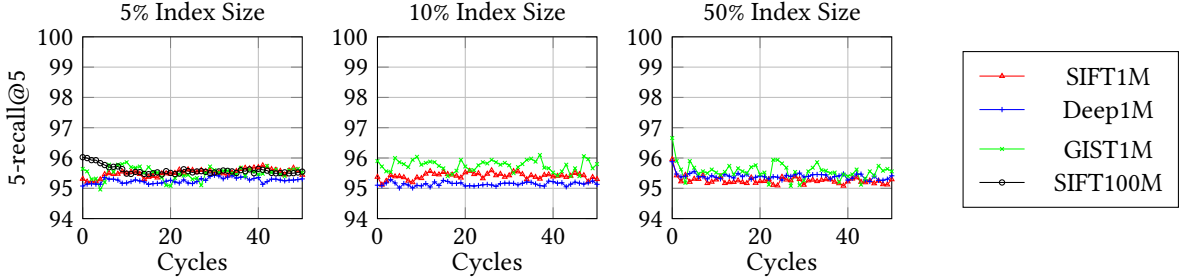
\begin{figure*}[ht]
  \begin{center}
  \begin{tikzpicture}
    \begin{groupplot}[group style={group size= 3 by 1}, height=4cm, width=4.5cm]
      \hspace{-30pt}
        \nextgroupplot[title=5\% Index Size, title style={yshift=-0.8ex}, ylabel={5-recall@5}, ylabel shift=-4pt,
        xlabel={Cycles}, xlabel shift=-4pt,ymin=94, ymax=100, xmin=0, xmax=50, grid=major, ytick distance = 1]
        \addplot [red, mark = triangle, mark size = 1pt]table[x=X,y=Y1,col sep=space] {plots/Lazy_reinc_5.txt}; \label{curve:sifts_5p}
        \addplot [blue, mark = +, mark size = 1pt]table[x=X,y=Y2,col sep=space] {plots/Lazy_reinc_5.txt}; \label{curve:deep_5p}
        \addplot [green, mark = x, mark size = 1pt]table[x=X,y=Y3,col sep=space] {plots/Lazy_reinc_5.txt}; \label{curve:gist_5p}
        \addplot [black, mark = o, mark size = 1pt]table[x=X,y=Y1,col sep=space] {plots/Lazy_reinc_sift100m_5.txt}; \label{curve:siftm_5p}
        \nextgroupplot[title=10\% Index Size, title style={yshift=-0.8ex}, ymin=94, ymax=100, xmin=0, xmax=50, xlabel={Cycles}, xlabel shift=-4pt,grid=major, ytick distance = 1]
        2       \addplot [red, mark = triangle, mark size = 1pt]table[x=X,y=Y1,col sep=space] {plots/Lazy_reinc_10.txt};
        \addplot [blue, mark = +, mark size = 1pt]table[x=X,y=Y2,col sep=space] {plots/Lazy_reinc_10.txt};
        \addplot [green, mark = x, mark size = 1pt]table[x=X,y=Y3,col sep=space] {plots/Lazy_reinc_10.txt};
        \nextgroupplot[title=50\% Index Size, title style={yshift=-0.8ex}, ymin=94, ymax=100, xmin=0, xmax=50, xlabel={Cycles}, xlabel shift=-4pt,grid=major, ytick distance = 1]
        2       \addplot [red, mark = triangle, mark size = 1pt]table[x=X,y=Y1,col sep=space] {plots/Lazy_reinc_50.txt};
        \addplot [blue, mark = +, mark size = 1pt]table[x=X,y=Y2,col sep=space] {plots/Lazy_reinc_50.txt};
        \addplot [green, mark = x, mark size = 1pt]table[x=X,y=Y3,col sep=space] {plots/Lazy_reinc_50.txt};
        \coordinate (top) at (rel axis cs:0,1);% coordinate at top of the first plot
        \coordinate (bot) at (rel axis cs:1,0);% coordinate at bottom of the last plot
    \end{groupplot}
    \path (top|-current bounding box.north) --
      coordinate(legendpos)
      (bot|-current bounding box.north);
\matrix[
    matrix of nodes,
    anchor=east,
    draw,
    inner sep=0.3em,
    column sep=0.5em,
    draw
  ]at([yshift=-12ex, xshift=35ex]legendpos)
  {
    \large{\ref{curve:sifts_5p}}& SIFT1M\\
    \large{\ref{curve:deep_5p}}& Deep1M\\
    \large{\ref{curve:gist_5p}}& GIST1M\\
    \large{\ref{curve:siftm_5p}}& SIFT100M\\};
    \end{tikzpicture}
  \vspace{-8pt}
  \caption{\recall{5}{5} for \algofresh indices for 50 cycles of deletion
    and re-insertion of 5\%, 10\%, and 50\% of index size on the million-point and 5\% of SIFT100M datasets. $L_s$ is chosen to obtain \recall{5}{5}$\approx 95\%$ on Cycle 0 index.}
  \label{fig:recall-stability-1M-100M}
\end{center}
\end{figure*}

\begin{comment}
\begin{figure}[t]
  \begin{center}
  \begin{tikzpicture}
  \begin{axis}[
            %title=Post-insertion recall,
            scale=0.42,
            grid=major, % Display a grid
            grid style={solid,gray!25}, % Set the style
            xmin = 0, xmax = 50,
            ymin = 94, ymax = 97,
            xlabel=\#Cycles (5\% size), % Set the labels
            xtick distance = 10,
            ylabel=5-recall@5,
            ylabel shift=-8pt,
            ytick distance = 1,
            legend columns = 1,
            legend style={at={(1.65,1)},anchor=north,font=\scriptsize},
          ]
          \addplot [red, mark = triangle, mark size = 0.5pt]table[x=X,y=Y1,col sep=space] {plots/Lazy_reinc_5.txt}; 
          \addplot [blue, mark = square, mark size = 0.5pt]table[x=X,y=Y2,col sep=space] {plots/Lazy_reinc_5.txt}; 
          \addplot [teal, mark = o, mark size = 0.5pt]table[x=X,y=Y3,col sep=space] {plots/Lazy_reinc_5.txt};
          \addplot [black, mark = x, mark size = 0.5pt]table[x=X,y=Y1,col sep=space] {plots/Lazy_reinc_sift100m_5.txt}; 
     \legend{SIFT1M L-20,DEEP1M L-21,GIST1M L-90}
  \end{axis}
  \end{tikzpicture}
  \vspace{-5pt}
  \caption{Search \recall{5}{5} of \algofresh index over 50 cycles of deletion
    and re-insertion of 5\% of index size on million-point datasets and SIFT100M. $L_s$ is chosen to obtain \recall{5}{5}$\approx 95\%$}
  \label{fig:recall-stability-1M-100M}
  \end{center}
  \vspace{-10pt}
\end{figure}
\end{comment}

We start with a statically built \algo index and subject it to multiple cycles
of insertions and deletions using the \algofresh update rules described
in~\Cref{sec:graph-updates}. In each cycle, we delete $5\%$, $10\%$ and $50\%$ of randomly
chosen points from the existing index, and re-insert the same points. 
We then choose appropriate $L_s$ (the
candidate list size during search) for $95\%$ 
\recall{5}{5} and plot the search recall as the index is updated. Since both the index contents and
$L_s$ are the same after each cycle, a good set of update rules would
keep the recall stable over these cycles.
~\Cref{fig:recall-stability-1M-100M} confirms that is indeed the case,
for the million point datasets and the 100 million point SIFT100M
dataset. In all these experiments, we use an identical set of parameters $L,\alpha, R$ for the static \algo index we begin with as well as our \algofresh updates. Note that in some of these plots, there is a small initial drop in recall; this is possibly due to the fact that the static \algo indices which we are starting from are built by making two passes of refinement over the dataset and hence might have slightly better quality than the streaming \algofresh algorithm. 

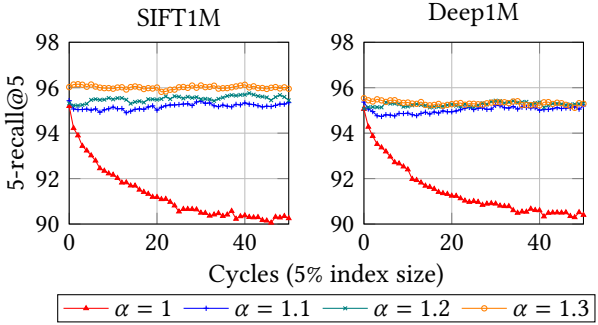
\begin{figure}[t]
  \begin{center}
  \begin{tikzpicture}
    \begin{groupplot}[group style={group size= 2 by 1}, height=4cm, width=4.5cm]
      \hspace{-10pt}
        \nextgroupplot[title=SIFT1M, title style={yshift=-0.8ex}, ylabel={5-recall@5}, ylabel shift=-4pt,ymin=90, ymax=98, xmax=50, xmin=0, grid=major]
        \addplot [red, mark = triangle*, mark size = 1pt]table[x=X,y=Y1,col sep=space] {appendix_plots/Alpha_vs_recall_build_updates_sift1m_ss5percent.txt}; \label{ss_alpha1}
				\addplot [blue, mark = +, mark size = 1pt]table[x=X,y=Y2,col sep=space] {appendix_plots/Alpha_vs_recall_build_updates_sift1m_ss5percent.txt}; \label{ss_alpha11}
				\addplot [teal, mark = x, mark size = 1pt]table[x=X,y=Y3,col sep=space] {appendix_plots/Alpha_vs_recall_build_updates_sift1m_ss5percent.txt}; \label{ss_alpha12}
				\addplot [orange, mark = o, mark size = 1pt]table[x=X,y=Y4,col sep=space] {appendix_plots/Alpha_vs_recall_build_updates_sift1m_ss5percent.txt}; \label{ss_alpha13}
        \coordinate (top) at (rel axis cs:0,1);% coordinate at top of the first plot
        \nextgroupplot[title=Deep1M, title style={yshift=-0.8ex}, ymin=90, ymax=98, xmin=0, xmax=50, grid=major]
        \addplot [red, mark = triangle*, mark size = 1pt]table[x=X,y=Y1,col sep=space] {appendix_plots/Alpha_vs_recall_build_updates_deep1m_ss5percent.txt};
				\addplot [blue, mark = +, mark size = 1pt]table[x=X,y=Y2,col sep=space] {appendix_plots/Alpha_vs_recall_build_updates_deep1m_ss5percent.txt};
				\addplot [teal, mark = x, mark size = 1pt]table[x=X,y=Y3,col sep=space] {appendix_plots/Alpha_vs_recall_build_updates_deep1m_ss5percent.txt};
				\addplot [orange, mark = o, mark size = 1pt]table[x=X,y=Y4,col sep=space] {appendix_plots/Alpha_vs_recall_build_updates_deep1m_ss5percent.txt};
        \coordinate (bot) at (rel axis cs:1,0);% coordinate at bottom of the last plot
    \end{groupplot}
    \path (top|-current bounding box.north) --
      coordinate(legendpos)
      (bot|-current bounding box.north);
\matrix[
    matrix of nodes,
    anchor=south,
    draw,
    inner sep=0.1em,
    column sep=0em,
    draw
  ]at([yshift=-29ex]legendpos)
  {
    \ref{ss_alpha1}& $\alpha=1$&[2pt]
    \ref{ss_alpha11}& $\alpha=1.1$&[2pt]
    \ref{ss_alpha12}& $\alpha=1.2$&[2pt]
    \ref{ss_alpha13}& $\alpha=1.3$\\};
\matrix[
    matrix of nodes,
    anchor=south,
    inner sep=0em,
  ]at([yshift=-26ex]legendpos)
  {Cycles (5\% index size)\\};
    \end{tikzpicture}
  \vspace{-8pt}
  \caption{Recall trends for \algofresh indices on SIFT1M and Deep1M
    over multiple cycles of inserting and deleting 5\% of points using
    different values of $\alpha$  for building and updating the
    index. $L_s$ is chosen to obtain $\recall{5}{5}\,\approx 95\%$ for
    Cycle 0 index.
  \vspace{-10pt}}
  \label{fig:alpha-ss-sift1m-5-recall}
\end{center}
\end{figure}

\smallskip \noindent {\bf Effect of $\alpha$.} Finally we study the effect of $\alpha$ on recall stability. In~\Cref{fig:alpha-ss-sift1m-5-recall}, we run the \algofresh update rules for a stream of deletions and insertions with different $\alpha$ values, and track how the recall changes as we perform our updates. Note that recall is stable for all indices except for the one with $\alpha = 1$,  validating the importance of using $\alpha > 1$. 

\section{The \diskanntwo system} \label{sec:final-design}

While \algofresh can support fast concurrent inserts, deletes and
searches with an in-memory index, it will not scale to a
billion-points per machine due to the large memory footprint of
storing the graph and data in RAM. The main idea of overall
system \diskanntwo is to store a bulk of the graph-index on an SSD,
and store only the recent changes in RAM.\footnote{As \algofresh
graphs are constructed using the $\alpha$-RNG property
(\Cref{sec:graph-updates}), the number of steps that the greedy search
algorithm takes to converge to a locally optima is much smaller than
other graph algorithms. Hence the total search latency to fetch
the graph neighborhoods from  SSD is small. So the $\alpha$-RNG
property helps us with both ensuring recall stability as well as
obtaining tolerable search latencies for SSD-based indices.} To
further reduce the memory footprint, we can simply
store \emph{compressed vector} representation (using an idea such as
Product Quantization (PQ)~\cite{PQ11}) of all the data vectors. In
fact, these ideas of using $\alpha$-RNG graphs and storing only
compressed vectors formed the crux of the SSD-based \diskann
static-ANNS index~\cite{DiskANN19}.
 
While this will reduce the memory footprint of our index, and will
also ensure reasonable search latencies, we cannot immediately run our
insert and delete~\Cref{alg:insert,alg:deletion} on to a
SSD-resident \algofresh index. Indeed, the insertion of a new point
$\dbpoint{p}$ has to update the neighborhoods of as many as $R$ (the
parameter controlling the degree bound) many points to add edges to
$p$, which would trigger up to $R$ random writes to the SSD. For
typical indices, $R$ would be as large as 64 or 128, requiring as many
random SSD writes per insert.  This would severely limit the insertion
throughput and also reduce the search throughput as a high write load
on the SSD also affects its read performance, which is critical to
search latency. Similarly, each delete operation, if applied eagerly,
would result in $R_{in}$ writes, where $R_{in}$ is the in-degree of
the deleted point, which can be very large.

The \diskanntwo system circumvents these issues and brings together
the efficiency of a SSD-based system and the interactive latency of
an in-memory system by \emph{splitting} the index into two parts: (i)
an in-memory \algofresh component comprising of recent updates, and
(ii) a larger SSD-resident index with longer term data.

\subsection{Components}

The overall system maintains two types of indices: one
\emph{Long-Term Index} (aka \lti) and one or more instances of
\emph{Temporary Index} (a.k.a \tempindex), along with a \deletelist.
\begin{itemize}
  \item \lti is an SSD-resident index that supports search
    requests. Its memory footprint is small, and consists only of
    about 25-32 bytes of compressed representations for each point. The
    associated graph index and full-precision data is stored on the
    SSD like~\cite{DiskANN19}. \emph{Insertions and deletions do not affect the \lti in
    real-time.}
    
  \item One or more \tempindex objects, which are instances of the
    \algofresh index stored entirely in DRAM (both the data and the
    associated graph). By design, they contain points that have been
    recently inserted to $P$. As a result, their memory footprint is a
    small fraction of the entire index.
  \item \deletelist is the list of points that are present either in the \lti or
    the \tempindex, but have been requested for deletion by the
    user. This list is used to filter out the deleted points returned in
    the search results.
\end{itemize}

{\bf RO- and RW-\tempindex}: To aid with crash recovery, \diskanntwo uses two types
of \tempindex. At all times, \diskanntwo will maintain one mutable
read-write \tempindex (called RW-\tempindex) which can accept insert requests. 
We periodically convert the RW-\tempindex into a \emph{read-only in-memory index} called
{RO-\tempindex}, and also snapshot it to persistent storage. 
We then create a new empty {RW-\tempindex} to ingest new points.

\subsection{\diskanntwo API}
The following three operations are supported:
\begin{itemize}
	\item \insertpt{\dbpoint{p}} to insert a new point to the index is
	routed to the sole instance of {RW-\tempindex}, which ingests the
	point using in~\Cref{alg:insert}.
	\item \deletept{p} request to delete an existing point $p$ is added
	to the \deletelist.
	\item \searchpt{\dbpoint{q}}{K}{L} to search for the $K$ nearest
	candidates using a candidate list of size $L$ is served by querying
	\lti, {RW-\tempindex}, and all instances of {RO-\tempindex} 
	with parameters $K$ and $L$, aggregating the results and removing
	deleted entries from \deletelist.
\end{itemize}

%Whenever triggered, the \merger
%merges all instances of the RO-\tempindex with the
%disk-resident \lti in the background (along with processing the relevant deletes in these indices marked in the \deletelist), and clears them out after completion.

%In the next subsection, we describe our \texttt{merge} procedure that takes in one \lti, one or more instances {RO-\tempindex} and a \deletelist and outputs a \emph{compact} \lti that folds inserted points from {RO-\tempindex} into \lti and removes references to vectors in \deletelist.

%\input{merge.tex}

\subsection{The \large{\merger} Procedure} \label{sec:ssd-updates}

Finally, to complete the system design, we now present details of the \merger procedure. Whenever the total memory footprint of the various {RO-\tempindex} exceeds a pre-specified threshold, the system invokes a background merge procedure serves to change the SSD-resident \lti to reflect the inserts from the various instances of the {RO-\tempindex} and also the deletes from the \deletelist. To this end, for notational convenience, let dataset $P$ reflect the points in the \lti, and $N$ denote points currently staged in the different RO-\tempindex instances, and $D$ denote the points marked for deletion in \deletelist. Then the desired end-result of the \merger is an SSD-resident \lti over
the dataset $(P \cup N) \setminus D$. Following the successful completion of the merge process, the system clears out the RO-\tempindex instances thereby keeping the total memory footprint under control. There are two important constraints
that the procedure must follow:
\begin{itemize}
\item Have a memory footprint proportional to size of the changes  $|D|$ and $|N|$, and not the size of overall index
  $|P|$. This is critical since the \lti can be much larger than
  the memory of the machine.
\item Use SSD I/Os efficiently so that searches can still be served
  while a merge runs in the background, and so that the merge itself
  can complete fast.
\end{itemize}

%% The input to the \merger process is a disk-resident \algofresh index
%% over a dataset $P$, the compressed vectors of each of the points in
%% $P$ (using the Product Quantization method~\cite{PQ11}), a set $N$ of
%% new points to insert into the index (which have been accumulated from
%% recent insertions into the fresh-ANNS index), and a set $D \subseteq
%% P$ of points to remove from $P$. The output is a new disk-resident
%% \algofresh index over the resulting set of points $(P \cup N)
%% \setminus D$.

%% We remark that
%% in the final system, this \merger procedure is run in the background
%% whenever $N$ becomes sufficiently large, and moreover, while it runs,
%% the overall system can still accept insert and delete operations in a
%% seamless manner to the end user. These mechanics are describe
%% in~\Cref{sec:final-design}.

At a high level, \merger first runs~\Cref{alg:deletion} to process the
deletes from $D$ to obtain an intermediate-\lti index over the points
$P \setminus D$. Then \merger runs~\Cref{alg:insert} to insert each of
the points in $N$ into the intermediate-\lti to obtain the resultant
\lti. However,~\Cref{alg:insert,alg:deletion} assume that both the
\lti graph, as well as the full-precision vectors all the datapoints
are stored in memory. The crucial challenges in \merger is to simulate
these algorithm invocations in a memory and SSD-efficient manner.
This is done in three phases outlined below.
\\[5pt]
%\begin{enumerate}
\noindent {\bf 1.}  \texttt{ Delete Phase:} This phase works on the
input \lti instance and produces an \emph{intermediate-\lti} by
running Algorithm ~\ref{alg:deletion} to process the deletions $D$.
To do this in a memory-efficient manner, we load the points in \lti
and their neighborhoods in the \lti \emph{block-by-block} from the
SSD, and execute Algorithm~\ref{alg:deletion} for the nodes in the
block using multiple threads, and write the modified block back to SSD
on the intermediate-\lti. Furthermore, whenever~\Cref{alg:deletion}
or~\Cref{alg:robustprune} make any distance comparisons, \emph{we use
  the compressed PQ vectors which are already stored on behalf of the
  \lti to calculate the approximate distances}. Note that this idea of
replacing any exact distance computations with approximate distances
using the compressed vectors will be used in the subsequent phases of
the \merger also.
\\[5pt]
\noindent {\bf 2.} \texttt{Insert Phase:} This phase adds all the new
points in $N$ to the intermediate-\lti by trying to simulate
Algorithm~\ref{alg:insert}.  As a first step, we run the
\greedysearch{s}{p}{1}{L} on the SSD-resident intermediate-\lti to get
the set $\cV$ of vertices visited on the search path.  Since the graph
is stored on the SSD, any requested neighborhood $\nout{p'}$ by the
search algorithm is fetched from the SSD. The $\alpha$-RNG property
ensures that the number of such neighborhood requests is small, and
hence the overall latency per point is bounded.  We then run the
\prune{p}{\cV}{\alpha}{R} procedure to determine the candidate set of
neighbors for $p$. However, unlike Algorithm~\ref{alg:insert}, we do
not immediately attempt to insert $p$ into $N_{out}(p')$ for $p'\in
N_{out}(p)$ (the backward edges) since this could result in an
impractical number of random reads and writes to the SSD. Instead, we
\emph{maintain an in-memory data-structure ${\bf \Delta}(p')$} and add
$p$ to that.
\\[5pt]
         %% Note that the size of ${\bf \Delta}$ is proportional to the
         %% set $N$ no matter how large the \lti is.
\noindent {\bf 3.}  \texttt{Patch Phase:} After processing all the
inserts, we patch the ${\bf \Delta}$ data-structure into the output
SSD-resident \lti index. For this, we fetch all points $p$ in the
intermediate-\lti block-by-block from the SSD, add the relevant
out-edges for each node $p$ from $\Delta$, and check the new degree
$|\nout{p} \cup {\bf \Delta}(p)|$ exceeds $R$. If so, prune the
neighborhood by setting $N_{out}(p) = \prune{p}{N_{out}(p)\cup {\bf
    \Delta}(p)}{\cdot}{\cdot}$. Within each block read from the SSD,
this operation can be applied to each vertex in a data-parallel
manner. Subsequently, the updated block is written back to SSD before
loading a new block.
%\end{enumerate}

\subsection{Complexity of \merger}
{\bf I/O cost}. The procedure does exactly two sequential passes over
the SSD-resident data structure in the Delete and Patch Phases. Due to
the $\alpha$-RNG property of the intermediate-\lti, the insertion
algorithm performs a small number of random 4KB reads per inserted
point (about 100 disk reads, a little more than the candidate list
size parameter, which we typically set to $75$). Note that this number
would be much larger without the $\alpha$-RNG property due to the
possibility of very long navigation paths.

{\bf Memory footprint:} Throughout the \merger process, ${\bf \Delta}$ data
structure has size $O(|N|R)$ where $R$ is the max-degree parameter 
of the index which is typically a small constant. For example,
if $|N| = 30M$ and $R = 64$, this footprint will be 
$\sim$7GB. In addition, for approximate distances, recall that we 
keep a copy of PQ coordinates for all points in the index ($\sim 32GB$ for a billion-point index).

{\bf Compute requirement:} The complexity of the insert phase and the
patch phase is essentially linear in the size of the new points $N$ to
insert, since the insert phase simply runs a search
using~\Cref{alg:greedysearch} for new point in $N$ and updates the
${\bf \Delta}$ data structure, and the patch phase adds the backward
edges in a block-by-block manner.

The delete phase has a small fixed cost to scan $\nout{p}$ of each
point $p\in P$ and check if there any deleted points and a larger
variable cost, linear in the delete set size $|D|$ that we will bound by $O(|D| R^2)$ (in expectation over random deletes).  We detail this calculation in~\Cref{app:compute-cost}.

\subsection{Recall Stability of \merger} \label{subsec:merger-stability}

While we have already demonstrated that our update algorithms~\Cref{alg:insert,alg:deletion} ensure recall stability over long streams of updates in~\Cref{subsec:algofresh_recall_stability}, the actual form in which these algorithms are implemented in our \merger procedure is different, especially with the use of approximate compressed vectors for distance computations. Indeed, as we process more cycles of the \merger
procedure, we expect the initial graph to be replaced by a graph
entirely built based on approximate distances. Hence, we
expect a small drop in recall in the initial cycles, following which we
expect the recall to stabilize. 

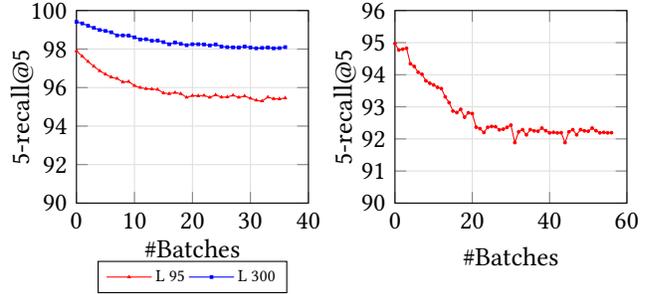
\begin{figure}[t]
  \begin{center}
    \begin{tikzpicture}
      \hspace{-0.3\columnwidth}
      \begin{axis}[
        scale=0.45,
        grid=major, % Display a grid
        grid style={solid,gray!25}, % Set the style
        xmin = 0, xmax = 40,
        ymin = 90, ymax = 100,
        xlabel=\#Batches, % Set the labels
        xlabel shift=-3pt,
        ylabel=\recall{5}{5},
        ylabel shift=-8pt,                   
        ytick distance = 2,
        legend columns = 2,
        legend style={at={(0.5, -0.3)},anchor=north, nodes={scale=0.65, transform shape}},
      ]
      \addplot [red, mark = triangle*, mark size = 0.5pt]table[x=X,y=Y1,col sep=space] {plots/SteadyState_sift100m_freshdiskann.txt}; 
      \addplot [blue, mark = square*, mark size = 0.5pt]table[x=X,y=Y2,col sep=space] {plots/SteadyState_sift100m_freshdiskann.txt}; 
  \legend{L 95,L 300}
  \end{axis}
\hspace{0.5\columnwidth}
	\begin{axis}[
		scale=0.45,
		grid=major, % Display a grid
		grid style={solid,gray!25}, % Set the style
		xmin = 0, xmax = 60,
		ymin = 90, ymax = 96,
		xlabel=\#Batches, % Set the labels
		ylabel= \recall{5}{5},
		ylabel shift=-5pt,                   
		ytick distance = 1,
		]
		\addplot [red, mark = *, mark size = 0.5pt]table[x=X,y=Y,col sep=space] {appendix_plots/SS_search_recall_800m_final.txt}; 
	\end{axis}
\end{tikzpicture}
%  \hspace{0.5\columnwidth}
%  %\begin{tikzpicture}
%    \begin{axis}[
%      scale=0.48,
%      grid=major, % Display a grid
%      grid style={solid,gray!25}, % Set the style
%      xmin = 0, xmax = 25,
%      ymin = 85, ymax = 100,
%      xlabel=SIFT1B ramp-up, % Set the labels
%      xlabel shift=-3pt,
%      ylabel=5-recall@5,
%      ylabel shift=-8pt,
%      ytick distance = 2,
%      legend columns = 3,
%      legend style={at={(0.5, -0.30)},anchor=north, nodes={scale=0.65, transform shape}},          
%    ]
%    \addplot [red, mark = triangle*, mark phase = 0, mark repeat = 1, mark size = 0.5pt]table[x=X,y=Y1,col sep=space] {plots/Ramp_up_sift1b_correct.txt}; 
%    \addplot [red]table[x=X,y=Y2,col sep=space] {plots/Ramp_up_sift1b_correct.txt}; 
%    \addplot [blue, mark = square*, mark phase = 0, mark repeat = 1, mark size = 0.5pt]table[x=X,y=Y3,col sep=space] {plots/Ramp_up_sift1b_correct.txt}; 
%    \addplot [blue]table[x=X,y=Y4,col sep=space] {plots/Ramp_up_sift1b_correct.txt}; 
%    \addplot [brown, mark = *, mark phase = 0, mark repeat = 1, mark size = 0.5pt]table[x=X,y=Y5,col sep=space] {plots/Ramp_up_sift1b_correct.txt}; 
%    \addplot [brown]table[x=X,y=Y6,col sep=space] {plots/Ramp_up_sift1b_correct.txt}; 
%  \legend{L 50, ,L 100, ,L 250}
%  \end{axis}
  %  \end{tikzpicture}
    \vspace{-5pt}
    \caption{Recall evolution over multiple cycles of \merger in
      \emph{steady-state} over (left) 80M point index with
      10\% deletes and inserts and (right) 800M point index with 30M insertions and deletions.}
  \label{fig:diskanntwo_recall_stab}
  \end{center}
  \vspace{-5pt}
\end{figure}

In the experiment in~\Cref{fig:diskanntwo_recall_stab}, we start with a statically built SSD-index built
on 80M points randomly sampled from the SIFT100M dataset. Then, in
each cycle, we update the index to reflect 8M deletions
and an equal number of insertions from the spare pool of 20M
points using \merger. We run this experiment for a total of 40 cycles and trace
recall for the index after each cycle
in~\Cref{fig:diskanntwo_recall_stab}. Note that the index stabilizes
at a lower recall value compared to the static index it
starts out with, due to the use of approximate distances in the
\merger process. We observe recall stabilization after $\approx 20$
cycles of deletion and insertion of 10\% of the index size, at which
point we expect most of the graph to be determined using approximate
distances. \Cref{fig:diskanntwo_recall_stab} (right)  shows a similar plot for the 800M point subset of SIFT1B. We have thus empirically demonstrated that the \diskanntwo index has stable recall over a stream of updates at steady-state.

%We now evaluate the recall stability of the \merger applied to
%SSD-resident \diskann indices in two settings -- \emph{steady-state}
%and \emph{ramp-up}. As the index evolves over multiple merge
%operations, we expect the following two factors to affect search
%recall for a fixed candidate list size $L$:
%\begin{itemize}
%	\item \textbf{Increasing index size.} In a \emph{ramp-up} setting, a
%	drop in recall is expected as the index grows in size and the search
%	becomes more complex.
	
%	\item \textbf{Approximate distance computation.} While the update
%	rules in \Cref{sec:graph-updates} use exact distance computations
%	for \algofresh indices, the \texttt{\merger} procedure uses Product
%	Quantization-based \emph{compressed data} to compute approximate
%	distances between points while determining the out-neighbors in
%	\texttt{Delete} and \texttt{Patch} phases. This could potentially
%	trade off graph quality for significantly reduced memory
%	footprint. Indeed, as we process more cycles of the \merger
%	procedure, we expect the initial graph to be replaced by a graph
%	entirely built based on approximate distances. Hence, we
%	expect a small drop in recall in the initial cycles, following which we
%	expect the recall to stabilize. Note that such a behavior would be
%	more pronounced when starting with a statically built graph.
%\end{itemize}

\subsection{Crash Recovery} \label{subsec:crash-recovery}

To support crash recover, all index update operations are written into a redo-log. When
a crash leads to the loss of the single {RW-\tempindex} instance and
the \deletelist, they are rebuilt by replaying updates from the
redo-log since the most recent snapshot. Since {RO-\tempindex}
and \lti instances are read-only and periodically snapshotted to disk, they can be simply reloaded from
disk.

The frequency at which {RW-\tempindex} is snapshotted to a
{RO-\tempindex} depends on the intended recovery time. More frequent
snapshots lead to small reconstruction times for {RW-\tempindex} but
create many instances of {RO-\tempindex} all of which have to be
searched for each query. While searching a few additional small
in-memory indices is not the rate limiting step for answering the
query (searching the large \lti is), creating too many could can lead
to inefficient search. A typical set up for a billion-point index would hold up to 30M points
in the \tempindex between merges to the \lti.  Limiting
each in-memory index to 5M points results in at most 6 instances
\tempindex which can each be searched in 0.77ms, compared to 0.89ms needed
to search a single 30M size index, for $Ls = 100$. On the flip side,
reconstructing the RW-\tempindex from the log using a 48 core machine takes just about 2.5
minutes if it has size 5M points as opposed to 16 minutes for a size of 30M points.

\section{Evaluation}\label{sec:eval}
We now study the \diskanntwo system on billion-scale datasets. We
first describe the datasets and the machines  used for all experiments reported in
this paper. We defer presentation of recall-vs-latency curves for \algofresh and \diskanntwo at $k=1, 10, 100$ to \Cref{sec:krecallk_appendix}.

%We now empirically demonstrate the recall stability of the \algofresh
%index and the \diskanntwo system over multiple cycles of insertions
%and deletions.  In fact, normalizing for index construction time, we
%show that the recall of the graph incrementally built by \algofresh is
%comparable to \algo, and better than popular algorithms like HNSW.
%Further, we measure the throughput and user facing latencies of the
%system in a realistic use case -- we incrementally build a (close-to)
%billion-point index, and run multiple cycles of deletions and
%insertions on this index over several days.

%% the graph update rules
%% in~\Cref{sec:graph-updates} keep the graph quality stable, i.e., the
%% same search parameter/complexity provides the same recall after index
%% modification (normalizing for index size).
%% Similarly, we demonstrate that the \diskanntwo system also has stable
%% recall over multiple rounds of modifications. Despite the streaming
%% merge procedure using approximate distances based on quantized data
%% instead of the full-precision data that the in-memory update algorithm
%% \algofresh uses, there is no excessive drop in graph quality.

\subsection{Experimental Setup}\label{subsec:exp-setup}
\noindent\textbf{Hardware}.  All experiments are run on one of two
machines:
\begin{itemize}
	\item (\texttt{mem-mc}) -- a 64-\emph{vcore} \texttt{E64d\_v4} Azure
	virtual machine instance used to measure latencies and recall for
	in-memory indices and the \algofresh update rules.
	\item (\texttt{ssd-mc}) -- a bare-metal server with 2x Xeon
          8160 CPUs (48 cores, 96 threads) and a 3.2TB Samsung PM1725a
          PCIe SSD to evaluate SSD-based indices and the overall \diskanntwo system.
\end{itemize}

\noindent\textbf{Datasets}. We evaluate our algorithms and systems on the following widely-used public benchmark datasets.
\begin{itemize}
\item 1 million point image descriptor datasets
  SIFT1M\cite{sift-link}, GIST1M\cite{sift-link}, and
  DEEP1M\cite{deep1b-link} in 128, 960 and 98 dimensions respectively.
  They are all in \texttt{float32}. DEEP1M is generated by
  convolutional neural networks.
\item 1 billion point SIFT1B\cite{sift-link} image descriptors in 128
  dimensions. It is the largest publicly available dataset and is in
  \texttt{uint8} precision (total data size 128GB). We take a random
  100M point subset of this dataset, represented in \texttt{float32}
  format and call it the SIFT100M dataset. We think that this smaller
  dataset captures many realistic medium-scale scenarios for ANNS.
\end{itemize}

%\input{expts-algofresh}
%\input{expts-ssdupdates}

%% \noindent\textbf{\algofresh}. Our experiments described 
%% in ~\Cref{subsec:algofresh_recall_stability} are run on the \texttt{mem}
%% machine with 64-threads for batch insert and delete operations. 
%% We present effective latencies for both insert and delete operations 
%% in ~\Cref{fig:ramp-up-100M}. Delete operations can be easily
%% handled by writing to a \emph{stop} list temporarily. However,
%% accumulating a large number of deletes in the stop list can degrade
%% search performance. Only for the purpose of this experiment, we
%% incorporate the time taken to reflect those changes in the \algofresh
%% index into the delete latency. As expected, with increasing index
%% size, both insert and delete operations take longer, but consistently
%% stay under 10ms per operation end-to-end.

%\input{plots/latency-ramp-up-1b}

\subsection{Billion-Scale \diskanntwo Evaluation}\label{subsec:ssd-perf}

We now study the complete \diskanntwo system in a realistic scenario
-- maintaining a large scale billion-scale index on the {\tt ssd}
machine and serving thousands of inserts, deletes and searches per
second concurrently over multiple days.  For this experiment, we use the SIFT1B dataset, but limit the size of
our indices to around 800M points, so that we have a sufficiently big
spare pool of 200M points for insertions at all times. 

\smallskip \noindent\textbf{Parameters}. We use $R=64, L_c = 75$ and
$\alpha=1.2$ for all the system. Recall that $R$ is the maximum degree of the graph, $L_c$
is the list size used during the candidate generation phase of the
algorithms (the parameter is used in~\Cref{alg:insert}), and $\alpha$
is used in the pruning phase for ensuring the $\alpha$-RNG property. We also use $B = 32$ bytes per data vector as the compression target in PQ (each data vector is compressed down to $32$ bytes) for the SSD-based \lti indices. We also set a limit $M$ of 30M points on the total size of the \tempindex so that the memory footprint of the \tempindex is bounded by around $13$GB ($128$ bytes per point for the vector data, $256$ bytes per point for the neighborhood information with $R = 64$, and some locks and other auxiliary data structures accounting for another $100$ bytes per point).  Finally,  we use a maximum of  $T = 40$  threads for  the \merger process which runs in the background.

\smallskip \noindent {\bf Memory Footprint of \diskanntwo Deployment.}
As mentioned above, the memory footprint of the \tempindex is around $13$ GB for 30M points, and our index will at any time store at most \tempindex instances totaling $60$M points, contributing a total of $\sim$26GB. The memory footprint index of the \lti for 800M points is essentially only the space needed to store the compressed vectors, which is around $24$ GB. The space requirement for the background \merger process is again at most $50$ GB (to store the compressed vectors of the 800M points of the \lti index and around $2\cdot R \cdot 4$ bytes per inserted point for forward and backward edges in the $\Delta$ data structure), giving us a peak memory footprint of around $100$GB. Since our index operated with a steady-state size of 800M points, this will roughly correspond to around $125$GB for a billion-point index. 

Our experiment can be divided into two phases: in the first phase, starting with a statically built index on a random 100M subset of SIFT1B, we define our \emph{update stream to comprise only of inserts} until the total number of points in the index reaches around 800M points. 
We  call this the ramp-up phase. We then transition into what we call a steady-state phase, where we update the index by deleting and inserting points at the same rate. We delete existing points and insert points from the spare pool of 200M points from the SIFT1B dataset. We then continue this for several days and observe the behaviour of the system in terms of latencies and recall.

How fast can we feed inserts into the system in these phases, i.e.,
how many threads can we use to concurrently insert into the
\diskanntwo system?  If we use too many threads for insertion, the
\tempindex will reach the limit $M$ of 30M points before the \merger
process has completed. This would result in a backlog of inserts not
consolidate to \lti on SSD. With the benefit of some prior experiments
(of how long each cycle of the \merger takes), we arrive at the number
of threads which concurrently feed inserts into the \diskanntwo system
in each of the phases and describe them below.

%Internally, we
%set a threshold of 30M points for merging in-memory data structures to
%the \lti . This ensures that the total RAM footprint of our index at
%all times including that of the \tempindex instances, delete list, the
%compressed data for \lti and any other buffers for \merger stays under
%$128$GB.

\medskip \noindent {\bf Stage 1: Ramp Up.}  In the first stage of the
experiment, we use the \diskanntwo system to start with an index of
100M points randomly chosen from the SIFT1B dataset, and constantly
feed inserts.  3 threads were used for concurrently inserting points
from the spare pool of points from SIFT1B, and 10 threads for issuing
concurrent search requests from the query set (with search parameters
set to provide $>92\%$ \recall{5}{5} at all times).  We chose $3$
threads for inserts so that the merge process does not get backlogged,
i.e., in the time taken by \merger to merge the previous batch of 30M
inserts to \lti, the \tempindex does not accumulate more than 30M
points. The insertions continued until the index grew to a size of
800M points, which took around 3 days. User-perceived mean search latency
over the course of the ramp-up fluctuates mostly between 5ms, when no
merge is happening, and 15ms when \merger is running in the background
and is presented in~\Cref{fig:ramp-up-1b-search-latency}.

\begin{figure}[t]
	\begin{center}
		\begin{tikzpicture}
			\begin{axis}[
				scale=0.7,
				grid=major, % Display a grid
				grid style={solid,gray!25}, % Set the style
				xmin = 0, xmax = 260000,
				ymin = 0, ymax = 40,
				xlabel=Time elapsed since beginning of experiment (seconds), % Set the labels
				xlabel shift =20 pt,
				ylabel=Search latency(ms),
				ylabel shift=-8pt,
				ytick distance = 5,        
				]
				\addplot [red, mark = triangle*, mark size = 0.5pt]table[x=X,y=Y,col sep=space] {Ramp_up_sift1b_concurr_search_latency.txt}; 
			\end{axis}
		\end{tikzpicture}
		\caption{Search latencies for $Ls=100$ (always $> 92\%$ \recall{5}{5}) over the course of ramping up an index to size 800M. Each point is mean latency over a 10000-query batch.}
		\label{fig:ramp-up-1b-search-latency}
	\end{center}
\end{figure}

%% The leftmost plot in \Cref{fig:steady-800m-insert-latency} shows

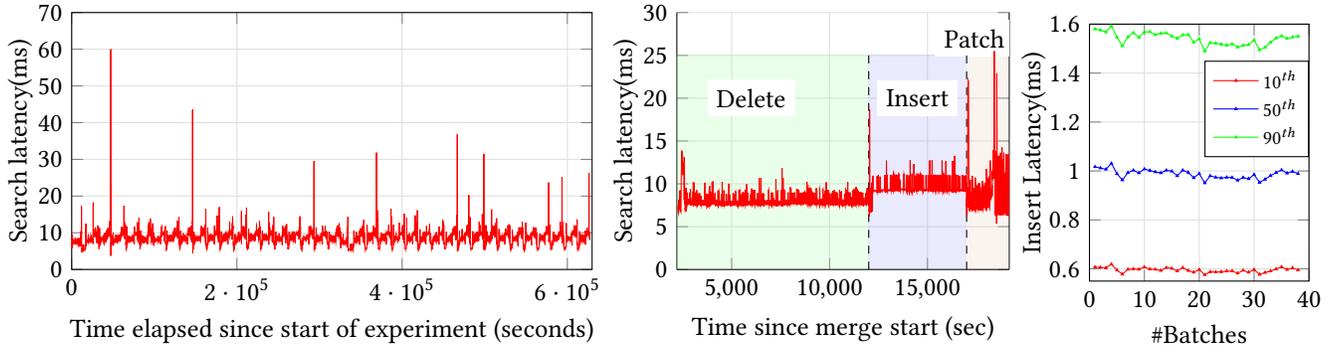
\begin{figure*}[t]
	%\begin{center}
		\begin{comment}
  \begin{tikzpicture}
	%\hspace{-0.75\columnwidth}
  \begin{axis}[
			width=6cm,
			height=5cm,
			grid=major, % Display a grid
			grid style={solid,gray!25}, % Set the style
			xmin = 0, xmax = 260000,
			ymin = 0, ymax = 40,
			xlabel=Time elapsed(seconds), % Set the labels
			xlabel shift =5 pt,
			ylabel=Search latency(ms),
			ylabel shift=-8pt,
			ytick distance = 10,    
			scaled x ticks=false    
		  ]
		  \addplot [red, mark = triangle*, mark size = 0.5pt]table[x=X,y=Y,col sep=space] {Ramp_up_sift1b_concurr_search_latency.txt}; 
  \end{axis}
  \end{tikzpicture}
\end{comment}
\begin{center}
\begin{tikzpicture}
	%\hspace{-0.4\columnwidth}
	\begin{axis}[
		width=8.5cm,
		height=5cm,
		grid=major, % Display a grid
		grid style={solid,gray!25}, % Set the style
		xmin = 0, xmax = 630000,
		ymin = 0, ymax = 70,
		xlabel= Time elapsed since start of experiment (seconds), % Set the labels
		ylabel=Search latency(ms),
		ylabel shift=-5pt,
		%xlabel shift =5pt,
		ytick distance = 10,
		xtick distance=200000,
		scaled x ticks=false,
		]
		\addplot [smooth, red, mark = none, mark size = 0.5pt]table[x=X,y=Y,col sep=space] {SteadyState_sift1b_concurr_search_latency.txt}; 		
	\end{axis}
%\end{tikzpicture}
%\hfill
\hspace{0.95\columnwidth}
  %\begin{tikzpicture}
  \begin{axis}[
		height=5cm,
		width=6cm,
		grid=major, % Display a grid
		grid style={solid,gray!25}, % Set the style
		axis x line*=bottom,
		scaled x ticks=false,
		xmin = 2200, xmax = 19173,
		ymin = 0, ymax = 30,
		xlabel= Time since merge start (sec), % Set the labels
		ylabel=Search latency(ms),
		ylabel shift=-5pt,
		ytick distance = 5
		]
		\filldraw[fill=green!40!white, fill opacity=0.2, draw=none] (0,0) rectangle (12000,25);
		\filldraw[fill=blue!40!white, fill opacity=0.2, draw=none] (12000,0) rectangle (17000, 25);
		\filldraw[fill=brown!40!white, fill opacity=0.2, draw=none] (17000,0) rectangle (20000, 25);
		\node[align=center, fill=white] at (6000, 20) {Delete};
		\node[align=center, fill=white] at (14500, 20) {Insert};
		\node[align=center, fill=white] at (17400, 27) {Patch};
		\draw[dashed] (12000, 0) -- (12000, 25);
		\draw[dashed] (17000, 0) -- (17000, 25);
		\addplot [smooth, red, mark = none, mark size = 0.5pt]table[x=X,y=Y1,col sep=space] {appendix_plots/SS_1b_40threads_search_latency.txt}; 
	\end{axis}
  \end{tikzpicture}
 % \caption{Search latencies for $Ls=100$ (always $> 92\%$ \recall{5}{5}) over the course of ramping up an index to size 800M. Each point is mean latency over a 10000-query batch.}
  %\label{fig:ramp-up-1b-search-latency}
  %\hspace{0.35\columnwidth}
	\hfill
		%\caption{Search latencies for 92\% \recall{5}{5} over a week-long steady-state \diskanntwo index processing concurrent inserts, deletes, and background merge on an 800M index. Each point is mean latency of a 10000-query batch.}
%		\label{fig:steady-800m-search-latency}
	%\hspace{0.75\columnwidth}
		\begin{tikzpicture}
			\begin{axis}[
				height=5cm,
				width=4.5cm,
				grid=major, % Display a grid
				grid style={solid,gray!25}, % Set the style
				xmin = 0, xmax = 40,
				ymin = 0.550, ymax = 1.600,
				xlabel=\#Batches, % Set the labels
				ylabel=Insert Latency(ms),
				ylabel shift=-6pt,                   
				ytick distance = 0.200,
				legend columns = 1,
				legend style={at={(1, 0.67)},anchor=east, nodes={scale=0.75, transform shape}},
				]
				\addplot [red, mark = triangle*, mark size = 0.5pt]table[x=X,y=Y4,col sep=space] {SteadyState_sift1b_insertion_latency_percentile.txt}; 
				\addplot [blue, mark = triangle*, mark size = 0.5pt]table[x=X,y=Y5,col sep=space] {SteadyState_sift1b_insertion_latency_percentile.txt}; 
				\addplot [green, mark = triangle*, mark size = 0.5pt]table[x=X,y=Y6,col sep=space] {SteadyState_sift1b_insertion_latency_percentile.txt}; 
				\legend{$10^{th}$, $50^{th}$, $90^{th}$}
			\end{axis}
		\end{tikzpicture}
		\vspace{-10pt}
		\caption{Mean latency\protect\footnotemark measurements for the week-long
                  \emph{steady-state} experiment with an 800M
                  \diskanntwo index processing concurrent inserts,
                  deletes, and periodic background merge.  (left)
                  Search latency with $Ls=100$ over the entire
                  experiment; (middle) Search latency during one
                  \merger run, zoomed in from the left plot; (right)
                  $10^{th}$, $50^{th}$ and $90^{th}$ percentile insert
                  latency over the entire experiment.}
		\label{fig:steady-800m-insert-latency}
	        \end{center}
                \vspace{-8pt}
\end{figure*}

\medskip \noindent {\bf Stage 2: Steady State.}  In the second stage
of the experiment, we maintain an index size of around 800M while
supporting a large number of equal inserts and deletes. Externally, 2
threads insert points into the index, 1 thread issues deletes, and 10
threads concurrently search it. Since the deletes happen near
instantly, we added a sleep timer between the delete requests to
ensure that the rate of deletions is similar to that of insertions.
Note that we reduced the number of insert threads from 3 to 2 to slow
down the insertion rate to accommodate the longer merge times compared
to the ramp-up experiment -- the \merger process now processes 30M
deletes in addition to 30M inserts. We present user-perceived
latencies for search and insertions
in~\Cref{fig:steady-800m-insert-latency}.

\medskip \noindent {\bf Variations in Search Latency During \merger.}
The middle plot in~\Cref{fig:steady-800m-insert-latency} shows that
the user-perceived search latencies varies across based on
the
phase of the \merger process in progress. Since the
\texttt{Insert} phase generates a significant number of random reads
to the \lti index which interfere with the random read requests issued
by the search threads,  it  results in slightly higher
latencies. On the other hand, while the typical latencies are smaller
during the \texttt{Delete} and \texttt{Patch} phases of
\texttt{\merger}, the latencies occasionally spike as high as 40ms,
which we think is likely due to head-of-line blocking by the large
sequential read and write operations that copy the \lti index to and
from the main memory.

%CAN THE SYSTEM CONTROL THESE RATES THAN THE USER FIGURING OUT?

% stacked bar chart instead of table
%\input{plots/merge-scaling-stacked-barchart.tex}

\begin{comment}
% old data in table format
\begin{table}
	\centering
	\caption{Merge time with different number of threads}
	\label{tab:merge-time-800M}
	\begin{tabular}{|c|c|c|c|c|} \hline
		\multirow{2}{*}{\#Threads} & Merge & Delete & Insert&Patch\\
									& Time(sec)&Phase&Phase&Phase\\ \hline
		10&56907 s&36230.7 s&13209.9 s&5534.19 s\\ \hline
		20&30047 s&18001.5 s&7187.45 s&2934.33s\\ \hline
		30&21387 s&12176.9 s&5260.63 s&2108.72s\\ \hline
		40&17119 s&9282.26 s&4290.84 s&1715.83s\\
		\hline\end{tabular}
\end{table}
\end{comment}

\smallskip \noindent{\bf Update Throughput of System.}
While \diskanntwo provides latencies of about $1$ms for insert
 (\Cref{fig:steady-800m-insert-latency}) and $0.1\mu $ for delete
 (since they are simply added to a \deletelist), in practice they need
 to be throttled so that the in-memory \tempindex do not grow too
 large before the ongoing background merge completes. As a result, the
 speed of the merge operation dictates the update rates the system can
 sustain over long periods of time. The threads allocation described
 above helps us control the rate of external insert and delete
 operations to what  the \texttt{\merger} procedure can
 complete before the \tempindex grows to $30$M points.

To better understand the thread allocation, we record the time taken
for the \merger process to merge 30M inserts into an index of size
roughly 800M using $T = 40$ threads. This takes around $8400$s per
cycle. To prevent the \tempindex from growing too much while the merge
procedure is running, we throttle the inserts to around $3500$ inserts
per second, so that the \tempindex accumulates under 30M newly
inserted points in one merge cycle. Since the insertion latencies into
in-memory \algofresh indices is around $1$ms
(\Cref{fig:steady-800m-insert-latency}), we allocated a total of $3$
threads concurrently feeding into the system. This ensured that the
system never backlogged throughout the ramp-up experiment.

%In the ramp-up experiment where the index grows without any delete
%operations, the merge takes a maximum of $8400$s to insert $30$M
%points into the \lti. We conclude that the system can support around
%$\frac{30,000,000}{8400}\approx3570$ inserts/second without
%backlogging the merge, when no delete requests are
%issued. We therefore used $3$ threads for insertions.

In the steady-state experiment where the index maintains a constant
size of about 800M points and is updated in cycles of equal sized
insertions and deletions of $30$M points, the
\texttt{\merger} procedure takes about 16277 seconds as it has to process deletes in addition to the inserts. Hence, in order to ensure that the system does not get backlogged, we throttled the insertion throughput to around $1800$ inserts per second (and similarly for deletes). We achieved this by using two threads for the insertions, and one thread (with a sleep timer) for deletes to match the insertion throughput.

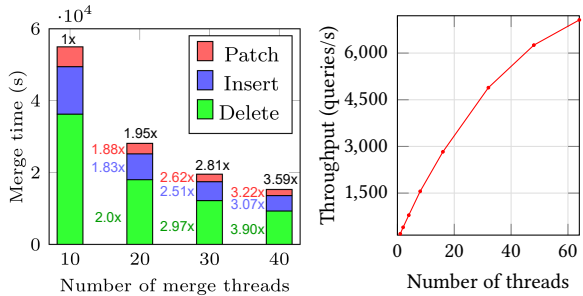
\begin{figure}[h]
	\vspace{-5pt}
	\includegraphics[width=4.0cm, height=3.9cm] {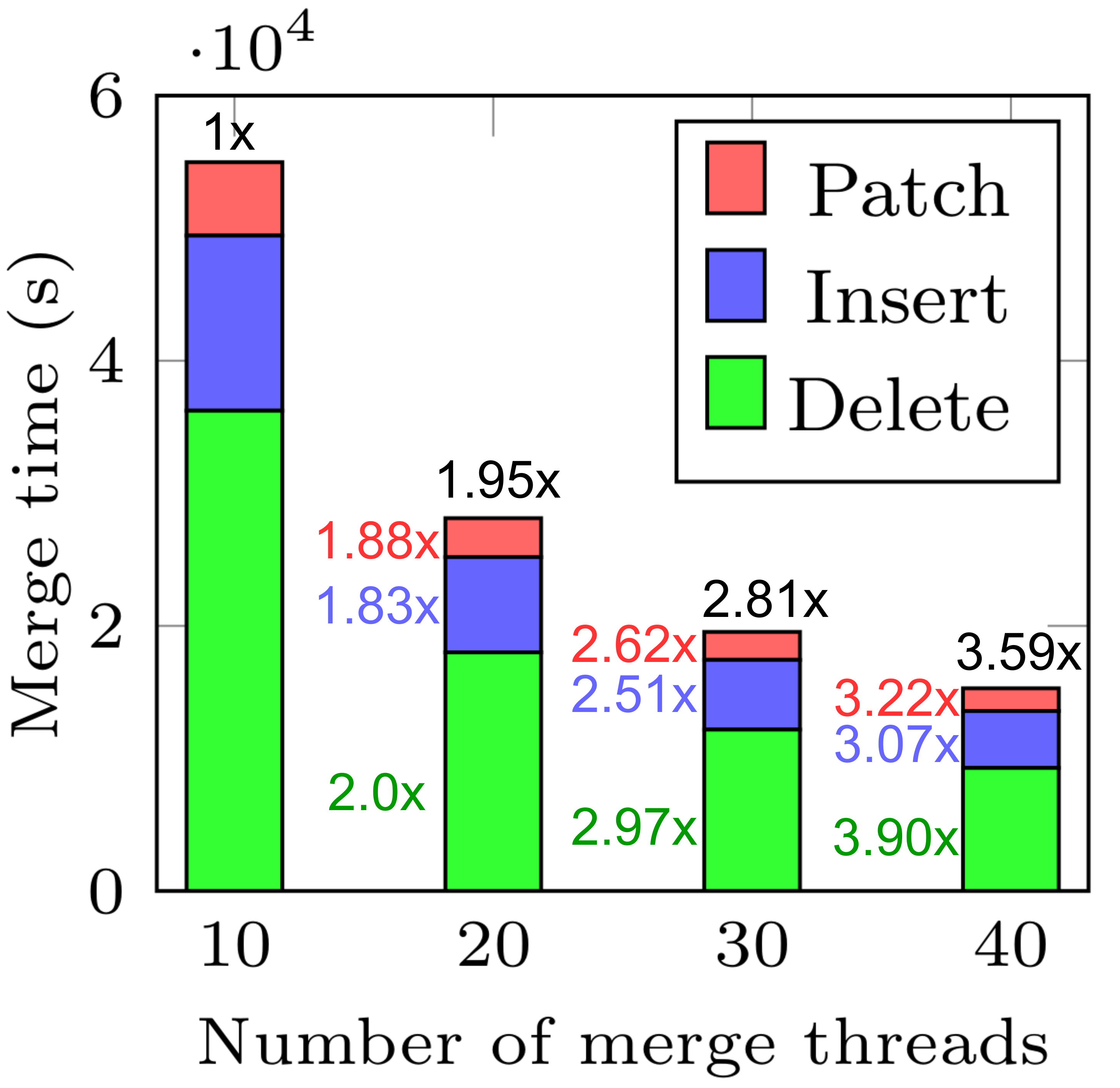}
	\hspace{-4pt}
	\begin{tikzpicture}
			\begin{axis}[
					grid=major, % Display a grid
					width=4cm,
					height=4.5cm,
					grid style={solid,gray!25}, % Set the style
					xmin = 0, xmax = 64,
					ymin = 150, ymax = 7200,
					xlabel= Number of threads, % Set the labels
					ylabel= Throughput (queries/s),
					x label style={font=\footnotesize},
					y label style={font=\footnotesize},
					xticklabel style={font=\footnotesize},
					yticklabel style={font=\footnotesize},
					ylabel shift=-5pt,
					xlabel shift = -1pt,
					ytick distance = 1500,
					]
					\addplot [red, mark = *, mark size = 0.5pt]table[x=X,y=Y,col sep=space] {SearchThroughput_800M.txt}; 
			\end{axis}
	\end{tikzpicture}

	\vspace{-5pt}
	\caption{(left) \texttt{\merger} runtime with different number of threads to merge 30M inserts and 30M deletes into a 800M SIFT index, and (right) Trend of search throughput with increasing search threads.}
	\vspace{-10pt}
	\label{fig:scaling-merge-search}
	\end{figure}

\begin{figure}[t]
	\begin{center}
		\begin{tikzpicture}
			\begin{axis}[
				scale=0.75,
				grid=major, % Display a grid
				grid style={solid,gray!25}, % Set the style
				axis x line*=bottom,
				xmin = 0, xmax = 34000,
				ymin = 0, ymax = 30,
				xlabel= Time elapsed(sec), % Set the labels
				ylabel=Search latency(ms),
				ylabel shift=-7pt,
				xlabel shift =20pt,
				ytick distance = 5
				]
				\addplot [smooth, red, mark = none, mark size = 0.5pt]table[x=X,y=Y1,col sep=space] {appendix_plots/SS_1b_20threads_search_latency2nd.txt}; 
%				\addplot [smooth, green, mark = none, mark size = 0.5pt]table[x=X,y=Y,col sep=space] {appendix_plots/Search_only_800m_10threads.txt}; 
			\end{axis}
			\begin{axis}[
				scale=0.75,
				grid=major, % Display a grid
				grid style={solid,gray!25}, % Set the style
				axis x line*=top,
				xmin = 0, xmax = 20000,
				ymin = 0, ymax = 30,
				ylabel=Search latency(ms),
				ylabel shift=-7pt,
				xlabel shift =20pt,
				ytick distance = 5
				]
				\filldraw[fill=green!40!white, fill opacity=0.2, draw=none] (2761,0) rectangle (12002,25);
				\filldraw[fill=blue!40!white, fill opacity=0.2, draw=none] (12002,0) rectangle (17054,25);
				\filldraw[fill=brown!40!white, fill opacity=0.2, draw=none] (17054,0) rectangle (18450,25);
				\node[align=center, fill=white] at (7000, 20) {Delete};
				\node[align=center, fill=white] at (14000, 20) {Insert};
				\node[align=center, fill=white] at (17700, 28) {Patch};
				\draw[dashed] (12002, 0) -- (12002, 25);
				\draw[dashed] (17054, 0) -- (17054, 25);
				\draw[dashed] (18450, 0) -- (18450, 25);
				\addplot [smooth, blue, mark = none, mark size = 0.5pt]table[x=X,y=Y1,col sep=space] {appendix_plots/SS_1b_40threads_search_latency.txt}; 				
			\end{axis}
		\end{tikzpicture}
		\caption{Trend of search latencies for 92\% search
			recall, zoomed in over one cycle of merging 30M inserts and deletes into a 800M index, using 20 threads (red) and 40 threads (blue) for merge (time-axes are normalized to align the phases).}
		\label{fig:steady-800m-fixed-search-threads}
	\end{center}
\end{figure}
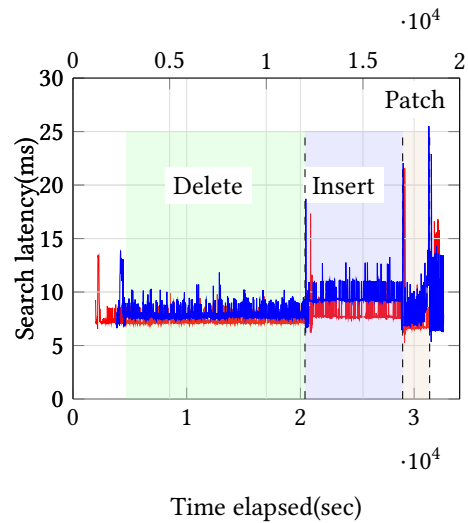

\smallskip \noindent{\bf Trade-off of Varying Number of Merge Threads $T$.} If we increase the merge threads $T$, the merges happen faster, which means we can ingest insertions and deletions into the system at a faster throughput (without the \tempindex size growing too large). On the other hand, if $T$ is large, the SSD-bandwidth used by the \merger process increases and this adversely affects the search throughput. We examine the merge times with varying threads in~\Cref{fig:scaling-merge-search} (left) and the search latencies when different numbers of threads are performing background merge in~\Cref{fig:steady-800m-fixed-search-threads}.

\smallskip \noindent{\bf I/O Cost of Search.}
Running search with candidate list size $L_s = 100$ gives us the
desired steady-state recall in these experiments. For this $L_s$
value, the average I/O complexity of searches ends up being a \emph{mere $120$ random $4$KB reads per query}, and the total number of distance comparisons made is around $8000$, \emph{a really tiny fraction of the cost of doing
brute force}. In contrast, systems like SRS~\cite{Sun14} end up
scanning $\approx 15\%$ of similar-sized datasets for achieving
moderate recall. 

\smallskip \noindent{\bf I/O Cost of Updates.}
Inserts and deletes involve reading and writing the
entire \lti ($\approx 320$GB), twice over. Since our system amortizes
this cost over $30M$ inserts and deletes, the SSD write cost per
update operation is around $10KB$, which is very small for a high
dimensional problem that requires data structure and algorithm with
random access patterns.
\footnotetext{Mean latency computed on a batch of 10k query points with one query per search thread}

\smallskip \noindent{\bf Scaling of Search Throughput.}
When the index is not processing any inserts, deletes or merges,
search throughput scales almost linearly with the number of threads
issuing search queries (see~\Cref{fig:scaling-merge-search}) (right), and with lesser latency than in
~\Cref{fig:steady-800m-insert-latency}. With 64 threads, the system
can support a throughput of $\sim 6500 $ queries/sec with a mean and
$99\%$ latency of under $10$ and $12$ms respectively. 

\smallskip \noindent{\bf The Cost of \merger.}
The \merger procedure with $40$ threads takes around
$16000$ seconds to merge 30M inserts and deletes into a 800M point \lti
(a 7.5\% change), which is $8.5\%$ of the $\approx 190000$
seconds it would take to re-build the index from scratch with a
similar thread-count. We conclude that the merge process is
significantly more cost-effective than periodically rebuilding the
indices, which is the current choice of system design for graph
indices. Further, \merger scales near linearly with the number of threads
(see~\Cref{fig:scaling-merge-search}). While the Delete phase scales linearly,
the Patch and Insert phases scale sub-linearly due to intensive SSD I/O.
Using fewer threads also results in more predictable search
latencies (esp. $99\%$ latency) due to the reduced SSD contention. 
This allows us to set the number of threads \merger uses to meet 
the desired update rate -- 3600 updates/sec require 40 threads, but
if we were only required to support 1000 updates/sec, we could choose
to run \merger with 10 threads, and take advantage of higher search
throughput and predictable latencies.

%% However, this comes at the cost of lower update rates due to the
%% need to limit the size of the \tempindex until merge completes. These
%% trade-offs are explored in~\Cref{fig:merge-scaling} and
%% ~\Cref{fig:steady-800m-search-latency-20threads}.

\begin{comment}
	\begin{figure}[h!]
		\begin{center}
			\includegraphics[width=0.67\columnwidth]{plots/merge_scalability_inv.pdf}
		\end{center}
		\caption{Runtime of \texttt{merger} with 10-40 threads}
		\label{fig:merge-scaling}
	\end{figure}
	\end{comment}

\section{Conclusion}
In this paper, we develop \algofresh, the first graph-based fresh-ANNS algorithm capable of reflecting updates to an existing index using compute proportional to the size of updates, while ensuring the index quality is similar to one rebuilt from scratch on the updated dataset. Using update rules from \algofresh, we design a novel two-pass \merger procedure which reflects these updates into an SSD-resident index with minimal write amplification. Using \algofresh and \merger, we develop and rigorously evaluate \diskanntwo, a highly-scalable fresh-ANNS system that can maintain a dynamic index of
 a billion points on a commodity machine while concurrently supporting inserts, deletes, and search operations at millisecond-scale latencies.
  
  \begin{comment}
While this represents a significant advance for large-scale ANNS, there are
shortcomings which present interesting research challenges.  Our
system only satisfies the somewhat weaker notion of quiescent
consistency. Can we extend our system to satisfy stronger consistency
properties such as linearizability? Another technically important problem is to reduce the variance in search latency during the different phases of the background merge process by
managing the SSD resource contention between the search and background processes.
  \end{comment}

\clearpage

\bibliographystyle{ACM-Reference-Format}
\bibliography{ref}

%%\newpage
\appendix

\section{Recall Stability of \algofresh Indices}

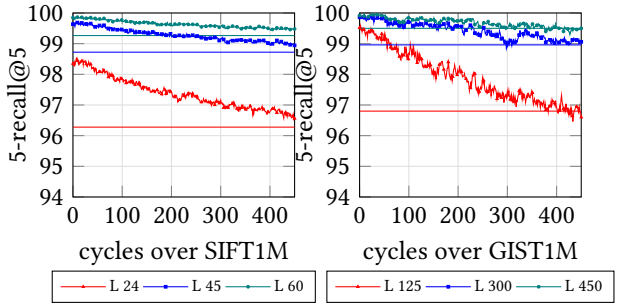
\begin{figure}[t]
%\begin{center}
  \begin{tikzpicture}
    \hspace{-0.25\columnwidth}
    \begin{axis}[
          %title=Recall on growing index after each cycle,
          scale=0.43,
          grid=major, % Display a grid
          grid style={solid,gray!25}, % Set the style
          xmin = 0, xmax = 450,
          ymin = 94, ymax = 100,
          xlabel=cycles over SIFT1M, % Set the labels
          ylabel=5-recall@5,
          ylabel shift=-8pt,
          xtick distance = 100,
          ytick distance = 1,
          legend columns = 3,
          legend style={at={(0.5,-0.4)},anchor=north,nodes={scale=0.65, transform shape}},
      ]
        \addplot [red, mark = triangle*, mark phase = 0, mark repeat = 9, mark size = 0.5pt]table[x=X,y=Y1,col sep=space] {plots/Ramp_up_sift1m.txt}; 
        \addplot [red]table[x=X,y=Y2,col sep=space] {plots/Ramp_up_sift1m.txt}; 
        \addplot [blue, mark = square*, mark phase = 0, mark repeat = 9, mark size = 0.5pt]table[x=X,y=Y3,col sep=space] {plots/Ramp_up_sift1m.txt}; 
        \addplot [blue]table[x=X,y=Y4,col sep=space] {plots/Ramp_up_sift1m.txt}; 
        \addplot [teal, mark = *, mark phase = 0, mark repeat = 9, mark size = 0.5pt]table[x=X,y=Y5,col sep=space] {plots/Ramp_up_sift1m.txt}; 
        \addplot [teal]table[x=X,y=Y6,col sep=space] {plots/Ramp_up_sift1m.txt}; 
        \legend{L 24, ,L 45, ,L 60, }
    \end{axis}
    \hspace{0.45\columnwidth}
    \begin{axis}[
          %title=Recall on growing index after each cycle,
          scale=0.43,
          grid=major, % Display a grid
          grid style={solid,gray!25}, % Set the style
          xmin = 0, xmax = 450,
          ymin = 94, ymax = 100,
          xlabel=cycles over GIST1M, % Set the labels
          ylabel=5-recall@5,
          ylabel shift=-8pt,
          xtick distance = 100,
          ytick distance = 1,
          legend columns = 3,
          legend style={at={(0.5,-0.4)},anchor=north,nodes={scale=0.65, transform shape}},
        ]
        \addplot [red, mark = triangle*, mark phase = 0, mark repeat = 9, mark size = 0.5pt]table[x=X,y=Y1,col sep=space] {plots/Ramp_up_gist1m.txt}; 
        \addplot [red]table[x=X,y=Y2,col sep=space] {plots/Ramp_up_gist1m.txt}; 
        \addplot [blue, mark = square*, mark phase = 0, mark repeat = 9, mark size = 0.5pt]table[x=X,y=Y3,col sep=space] {plots/Ramp_up_gist1m.txt}; 
        \addplot [blue]table[x=X,y=Y4,col sep=space] {plots/Ramp_up_gist1m.txt}; 
        \addplot [teal, mark = *, mark phase = 0, mark repeat = 9, mark size = 0.5pt]table[x=X,y=Y5,col sep=space] {plots/Ramp_up_gist1m.txt}; 
        \addplot [teal]table[x=X,y=Y6,col sep=space] {plots/Ramp_up_gist1m.txt}; 
        \legend{L 125, ,L 300, ,L 450, }
    \end{axis}
  \end{tikzpicture}
  \vspace{-6pt}
  \caption{Search \recall{5}{5} after each cycle of 12K insertions and
    10K deletions to \algofresh, ramping up from 100K to 1M
    points. Horizontal lines indicate recall of the corresponding
    batch built \algo index.}
  \label{fig:ramp-up-1M}
  \vspace{-10pt}
%\end{center}
\end{figure}

\smallskip \noindent {\bf Ramp-Up.} We now measure the recall of an
\algofresh index as it grows in size. We start with a \algo index
built on a subset of 100K points randomly sampled from the million
point datasets. In each cycle, we delete 10K points from the index at
random, and insert 12K new points from the remaining pool of points,
so that index grows by 2000 points. The experiment concludes when the
index reaches the full size of a million points.  We plot the search
recall varies over the cycles in \Cref{fig:ramp-up-1M} with varying
search list size.  While the recall trends down for a fixed search
list size $L$ as expected\footnote{This is true of any index -- a
	larger index over data from the same distribution will provide lower
	recall with the search parameter/complexity.}, note that the
	\emph{final index quality} is at least as good as indices built in one
	shot using \algo, whose recall for the same parameters is marked by
	horizontal lines.
	
	\begin{figure}[t]
\begin{center}
  \begin{tikzpicture}
    \hspace{-0.25\columnwidth}
\begin{axis}[
          %title=Recall on growing index after each cycle,
          scale=0.44,
          grid=major, % Display a grid
          grid style={solid,gray!25}, % Set the style
          xmin = 0, xmax = 70,
          ymin = 93, ymax = 100,
          xlabel= \#cycles, % Set the labels
          ylabel=5-recall@5,
          ylabel shift=-8pt,
          ytick distance = 1,
          legend columns = 3,
          legend style={at={(0.5,-0.4)},anchor=north,nodes={scale=0.65, transform shape}},
  ]
        \addplot [red, mark = triangle*, mark size = 0.5pt]table[x=X,y=Y1,col sep=space] {Full_ramp_up_sift100m_2.txt}; 
        \addplot [red]table[x=X,y=Y2,col sep=space] {Full_ramp_up_sift100m_2.txt}; 
        \addplot [blue, mark = square*, mark size = 0.5pt]table[x=X,y=Y3,col sep=space] {Full_ramp_up_sift100m_2.txt}; 
        \addplot [blue]table[x=X,y=Y4,col sep=space] {Full_ramp_up_sift100m_2.txt}; 
        \addplot [teal, mark = *, mark size = 0.5pt]table[x=X,y=Y5,col sep=space] {Full_ramp_up_sift100m_2.txt}; 
        \addplot [teal]table[x=X,y=Y6,col sep=space] {Full_ramp_up_sift100m_2.txt}; 
\legend{L 60, ,L 150, ,L 220}
\end{axis}
%\end{tikzpicture}
%\caption{Search recall over the course of ramping up a static index on a single point from SIFT1B upto 100M, stepping up by 1.5M points in every batch}
%\label{fig:fig13}
%\end{center}
%\end{figure}
\hspace{0.45\columnwidth}
%\begin{figure}[t]
%\begin{center}
%\begin{tikzpicture}
\begin{axis}[
          %title=Recall on growing index after each cycle,
          scale=0.44,
          grid=major, % Display a grid
          grid style={solid,gray!25}, % Set the style
          xmin = 0, xmax = 55,
          ymin = 93, ymax = 100,
          xlabel=\#cycles, % Set the labels
          ylabel=5-recall@5,
          ylabel shift=-8pt,
          ytick distance = 1,
          legend columns = 3,
          legend style={at={(0.5,-0.4)},anchor=north,nodes={scale=0.65, transform shape}},
        ]
        \addplot [red, mark = triangle*, mark size = 0.5pt]table[x=X,y=Y1,col sep=space] {plots/Ramp_up_sift100m.txt}; 
        \addplot [red]table[x=X,y=Y2,col sep=space] {plots/Ramp_up_sift100m.txt}; 
        \addplot [blue, mark = square*, mark size = 0.5pt]table[x=X,y=Y3,col sep=space] {plots/Ramp_up_sift100m.txt}; 
        \addplot [blue]table[x=X,y=Y4,col sep=space] {plots/Ramp_up_sift100m.txt}; 
        \addplot [teal, mark = *, mark size = 0.5pt]table[x=X,y=Y5,col sep=space] {plots/Ramp_up_sift100m.txt}; 
        \addplot [teal]table[x=X,y=Y6,col sep=space] {plots/Ramp_up_sift100m.txt}; 
\legend{L 55, ,L 100, ,L 200}
\end{axis}
  \end{tikzpicture}
  \vspace{-5pt}
  \caption{ Search recall \algofresh on SIFT100M while (left) ramping
    up from 1 point; and (right) ramping up starting from 30M points,
    and steady-state after $45$ cycles. Horizontal lines indicate
    recall of the \algo index with the same build time.}
  \vspace{-12pt}
\label{fig:ramp-up-100M}
\end{center}
\end{figure}
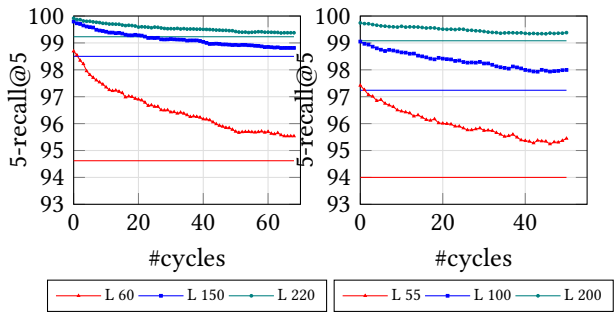

\section {Index build times}\label{sec:build-times}
In ~\Cref{tab:build-time} we compare the build time of \algo and \algofresh for the same build parameters.
The trade-off for this speed-up comes in the form of increased search latency for the same \recall{k}{k}. 
In ~\Cref{fig:norm-rebuild-chart}, we show that using \algofresh to make updates to the index takes only a fraction of the time to rebuild it from scratch using \algo.
We show a similar comparison of \diskann and \diskanntwo in ~\Cref{tab:diskann-update-time-30m}. 
Despite using more than double the resources, building a 800M index from scratch using \diskann takes more than 7x the time that \diskanntwo takes to reflect the same changes into the index.
\begin{table}[t]
	\centering
	\caption{Index build times for \algo and \algofresh on \texttt{mem} with $R=64, L_c=75, \alpha=1.2$}
	\label{tab:build-time}
	\begin{tabular}{|c|c|c|c|} \hline
		Dataset&\algo & {\algofresh}& Speedup\\ \hline
		SIFT1M&32.3s&21.8 s & 1.48x \\ \hline
		DEEP1M&26.9s&17.7 s & 1.52x \\ \hline
		GIST1M&417.2s&228.1 s & 1.83x \\ \hline
		SIFT100M&7187.1s&4672.1s & 1.54x \\
		\hline\end{tabular}
\end{table}

\pgfplotsset{compat=newest}
\begin{figure}[t]
\begin{center}
\begin{tikzpicture}
\pgfplotstableread{
0.116563467 
0.218894737
0.623585139
0.118717472
0.241847584
0.595821561
0.058349712
0.131045781
0.523701103
0.101393191
0.189770561
0.544718871
}\datatable

\node [align=left, font=\small,
    text width=1.5cm, inner sep=0.25cm] at (1.18cm, -0.90cm) {\textsc{SIFT1M}};

\node [align=left, font=\small,
text width=1.5cm, inner sep=0.25cm] at (2.75cm, -0.90cm) {\textsc{DEEP1M}};

\node [align=left, font=\small,
text width=1.5cm, inner sep=0.25cm] at (4.32cm, -0.90cm) {\textsc{GIST1M}};

\node [align=left, font=\small,
text width=1.5cm, inner sep=0.25cm] at (5.9cm, -0.90cm) {\textsc{SIFT100M}};

\begin{axis}[
   ybar,
    title style={align=center},
    ticks=both,
    ytick={0.1, 0.2, 0.5, 1.0},
    axis x line = bottom,
    axis y line = left,
    axis line style={-|},
    %nodes near coords = \rotatebox{90}{{\pgfmathprintnumber[fixed zerofill, precision=2]{\pgfplotspointmeta}}},
    nodes near coords align={vertical},
    point meta={y*100},
    nodes near coords=\rotatebox{90}{\pgfmathprintnumber[fixed zerofill, precision=1]\pgfplotspointmeta\%},
    every node near coord/.append style={font=\footnotesize, fill=none, yshift=0.5mm},
    enlarge y limits={lower, value=0.1},
    enlarge y limits={upper, value=0.22},
    ylabel=Relative Merge time,
   xtick=data,
   /pgf/bar width=6pt,
    ymin = 0,
    ymajorgrids,
    xticklabels={ 
        5\%, 
        10\%,
        50\%,
        5\%, 
        10\%,
        50\%,
        5\%, 
        10\%,
        50\%,
        5\%, 
        10\%,
        50\%},
    legend style={at={(0.5, -0.30)}, anchor=north, legend columns=3},
    every axis legend/.append style={nodes={right}, inner sep = 0.1cm},
   x tick label style={rotate=45, align=center, yshift=0cm, font=\footnotesize},
   y tick label style={xshift=0.0cm, font=\small},
    enlarge x limits=0.1,
    width=0.47\textwidth,
    height=5cm,
    ]
\pgfplotsinvokeforeach {0,...,0}{
    \addplot table [x expr={\coordindex-mod(\coordindex, 3)/6}, y index=#1] {\datatable};
}

\draw (axis cs:2.33,0) -- ({axis cs:2.33,0}|-{rel axis cs:0.5,1});
\draw (axis cs:5.33,0) -- ({axis cs:5.33,0}|-{rel axis cs:0.5,1});
\draw (axis cs:8.33,0) -- ({axis cs:8.33,0}|-{rel axis cs:0.5,1});
% \draw (axis cs:11.33,0) -- ({axis cs:11.33,0}|-{rel axis cs:0.5,1});
%\legend{\algofresh \hspace*{8pt}, HNSW} 
\end{axis}
\end{tikzpicture}
\end{center}
\caption{Time taken to merge delete and re-insert of 5\%, 10\%, and 50\% of index size into a \algofresh index, expressed relative to index rebuild time for \algo.}
\label{fig:norm-rebuild-chart}
\end{figure}
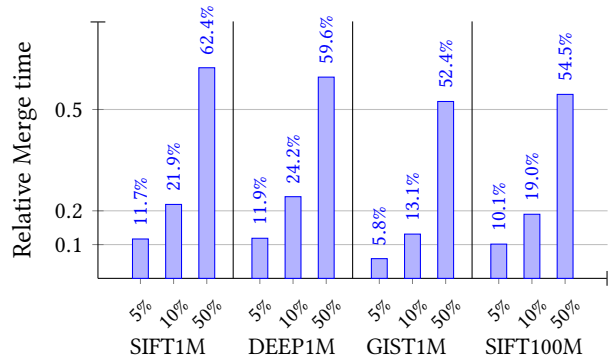

\begin{table}
	\centering
	\caption{Full build time with \diskann ($96$ threads) versus \diskanntwo ($40$ threads) to update a 800M index with 30M inserts and deletes}
	\label{tab:diskann-update-time-30m}
	\begin{tabular}{|c|c|c|} \hline
		Dataset&DiskANN(sec)&\merger(sec)\\ \hline
		SIFT800M&83140 s&15832 s\\
		\hline\end{tabular}
\end{table}

\section{Effect of $\alpha$ on recall stability}
\begin{figure}[ht]
  \begin{center}
  \begin{tikzpicture}
    \begin{groupplot}[group style={group size= 2 by 1}, height=4cm, width=4.5cm]
      \hspace{-10pt}
        \nextgroupplot[title=SIFT1M, title style={yshift=-0.8ex}, ylabel={5-recall@5}, ylabel shift=-4pt,ymin=95, ymax=100, xmin=100, xmax=600, grid=major, ytick distance = 1]
        \addplot [red, mark = triangle*, mark size = 1pt]table[x=X1,y=Y1,col sep=space] {appendix_plots/Alpha_recall_vs_latency_sift1m.txt}; \label{ss_alpha1}
				\addplot [blue, mark = +, mark size = 1pt]table[x=X2,y=Y2,col sep=space] {appendix_plots/Alpha_recall_vs_latency_sift1m.txt}; \label{ss_alpha11}
				\addplot [teal, mark = x, mark size = 1pt]table[x=X3,y=Y3,col sep=space] {appendix_plots/Alpha_recall_vs_latency_sift1m.txt}; \label{ss_alpha12}
				\addplot [orange, mark = o, mark size = 1pt]table[x=X4,y=Y4,col sep=space] {appendix_plots/Alpha_recall_vs_latency_sift1m.txt}; \label{ss_alpha13}
        \coordinate (top) at (rel axis cs:0,1);% coordinate at top of the first plot
        \nextgroupplot[title=Deep1M, title style={yshift=-0.8ex}, ymin=95, ymax=100, xmin=80, xmax=500, grid=major, ytick distance = 1]
        \addplot [red, mark = triangle*, mark size = 1pt]table[x=X1,y=Y1,col sep=space] {appendix_plots/Alpha_recall_vs_latency_deep1m.txt};
				\addplot [blue, mark = +, mark size = 1pt]table[x=X2,y=Y2,col sep=space] {appendix_plots/Alpha_recall_vs_latency_deep1m.txt};
				\addplot [teal, mark = x, mark size = 1pt]table[x=X3,y=Y3,col sep=space] {appendix_plots/Alpha_recall_vs_latency_deep1m.txt};
				\addplot [orange, mark = o, mark size = 1pt]table[x=X4,y=Y4,col sep=space] {appendix_plots/Alpha_recall_vs_latency_deep1m.txt};
        \coordinate (bot) at (rel axis cs:1,0);% coordinate at bottom of the last plot
    \end{groupplot}
    \path (top|-current bounding box.north) --
      coordinate(legendpos)
      (bot|-current bounding box.north);
\matrix[
    matrix of nodes,
    anchor=south,
    draw,
    inner sep=0.1em,
    column sep=0em,
    draw
  ]at([yshift=-29ex]legendpos)
  {
    \ref{ss_alpha1}& $\alpha=1$&[2pt]
    \ref{ss_alpha11}& $\alpha=1.1$&[2pt]
    \ref{ss_alpha12}& $\alpha=1.2$&[2pt]
    \ref{ss_alpha13}& $\alpha=1.3$\\};
\matrix[
    matrix of nodes,
    anchor=south,
    inner sep=0em,
  ]at([yshift=-26ex]legendpos)
  {Mean query latency ($\mu$s)\\};
    \end{tikzpicture}
  \vspace{-8pt}
  \caption{Recall vs latency curves for \algo indices on SIFT1M and Deep1M built with different values of $\alpha$\vspace{-10pt}}
  \label{fig:alpha-ss-latency-recall}
\end{center}
\end{figure}
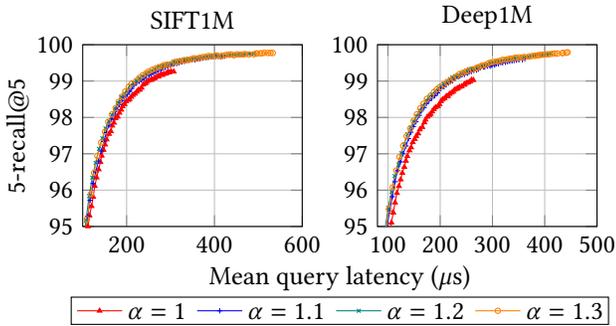

To determine the optimal value of $\alpha$, we perform the \algofresh steady-state experiments with different values of $\alpha$.
In the plots in~\Cref{fig:alpha-ss-sift1m-5-recall}, we use the same value of $\alpha$ for building the intial \algo index and for updating it. 
Other build and update parameters are same for each plot (R = 64, L = 75).
We compare the evolution of search recall in the $95\%$ range and average degree with different $\alpha$. 
Finally we compare search recall versus latency for static indices built with different $\alpha$ to choose the best candidate.
For all $\alpha > 1$, average degree increases over the course of the experiments and recall stabilises around the initial value.
For static indices, latency at the same recall value improves from 1 to 1.2 after which further increasing $\alpha$ shows now noticeable improvement as evidenced by recall-vs-latency plots for \algo indices in \Cref{fig:alpha-ss-latency-recall}.
Since we want to minimise the memory footprint of our index, we choose the $\alpha$ value with best search performance and lowest average degree, which in this case is 1.2.
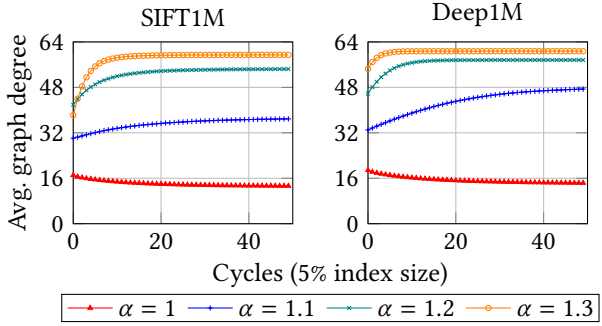
\begin{figure}[ht]
  \begin{center}
  \begin{tikzpicture}
    \begin{groupplot}[group style={group size= 2 by 1}, height=4cm, width=4.5cm]
      \hspace{-10pt}
        \nextgroupplot[title=SIFT1M, title style={yshift=-0.8ex}, ylabel={Avg. graph degree}, ylabel shift=-4pt,ymin=0, ymax=64, xmax=50, xmin=0, grid=major, ytick distance = 16]
        \addplot [red, mark = triangle*, mark size = 1pt]table[x=X,y=Y1,col sep=space] {appendix_plots/Alpha_vs_degree_build_updates_sift1m_ss5percent.txt}; \label{ss_alpha1}
				\addplot [blue, mark = +, mark size = 1pt]table[x=X,y=Y2,col sep=space] {appendix_plots/Alpha_vs_degree_build_updates_sift1m_ss5percent.txt}; \label{ss_alpha11}
				\addplot [teal, mark = x, mark size = 1pt]table[x=X,y=Y3,col sep=space] {appendix_plots/Alpha_vs_degree_build_updates_sift1m_ss5percent.txt}; \label{ss_alpha12}
				\addplot [orange, mark = o, mark size = 1pt]table[x=X,y=Y4,col sep=space] {appendix_plots/Alpha_vs_degree_build_updates_sift1m_ss5percent.txt}; \label{ss_alpha13}
        \coordinate (top) at (rel axis cs:0,1);% coordinate at top of the first plot
        \nextgroupplot[title=Deep1M, title style={yshift=-0.8ex}, ymin=0, ymax=64, xmin=0, xmax=50, grid=major, ytick distance = 16]
        \addplot [red, mark = triangle*, mark size = 1pt]table[x=X,y=Y1,col sep=space] {appendix_plots/Alpha_vs_degree_build_updates_deep1m_ss5percent.txt};
				\addplot [blue, mark = +, mark size = 1pt]table[x=X,y=Y2,col sep=space] {appendix_plots/Alpha_vs_degree_build_updates_deep1m_ss5percent.txt};
				\addplot [teal, mark = x, mark size = 1pt]table[x=X,y=Y3,col sep=space] {appendix_plots/Alpha_vs_degree_build_updates_deep1m_ss5percent.txt};
				\addplot [orange, mark = o, mark size = 1pt]table[x=X,y=Y4,col sep=space] {appendix_plots/Alpha_vs_degree_build_updates_deep1m_ss5percent.txt};
        \coordinate (bot) at (rel axis cs:1,0);% coordinate at bottom of the last plot
    \end{groupplot}
    \path (top|-current bounding box.north) --
      coordinate(legendpos)
      (bot|-current bounding box.north);
\matrix[
    matrix of nodes,
    anchor=south,
    draw,
    inner sep=0.1em,
    column sep=0em,
    draw
  ]at([yshift=-29ex]legendpos)
  {
    \ref{ss_alpha1}& $\alpha=1$&[2pt]
    \ref{ss_alpha11}& $\alpha=1.1$&[2pt]
    \ref{ss_alpha12}& $\alpha=1.2$&[2pt]
    \ref{ss_alpha13}& $\alpha=1.3$\\};
\matrix[
    matrix of nodes,
    anchor=south,
    inner sep=0em,
  ]at([yshift=-26ex]legendpos)
  {Cycles (5\% index size)\\};
    \end{tikzpicture}
  \vspace{-8pt}
  \caption{Evolution trends of recall and average degree for \algofresh indices on SIFT1M and Deep1M over multiple cycles of inserting and deleting 5\% of the index, where each trend represents a different $\alpha$ value used for building and updating the index.
  \vspace{-10pt}}
  \label{fig:alpha-ss-sift1m-5-degree}
\end{center}
\end{figure}

\section{Amortized Cost of Delete Phase in \merger} \label{app:compute-cost}

Any non-trivial computation in the delete phase happens {only for undeleted points $p \in P$
	which have neighbors from the delete list}. For each such point
$p$,~\Cref{alg:deletion} applies the pruning process on the candidate
list consisting of the undeleted neighbors of $p$ and the undeleted
neighbors of the deleted neighbors of $p$ to select the best $R$
points from to its updated neighborhood. In order to perform an
average-case analysis, let us assume that the delete set $D$ is a
randomly chosen set from the active points $P$, and suppose $|P| = N$
and $\frac{|D|}{N} = \beta$.  The expected size of the candidate list
will be $R(1-\beta) + {R^2}\beta(1-\beta)$.  Here the first term
accounts for undeleted neighbors of $p$ and the second term accounts
for undeleted neighbors of deleted neighbors of $p$. The expected
number of undeleted points in the index is $N(1-\beta)$. Therefore the
total number of expected operations in the delete phase will be
proportional to $NR{(1-\beta)}^2(1+R\beta)$. This assumes that the
complexity of the prune procedure is linear in the size of the
candidate list which we validated empirically below.  For large values of
$\beta$, the ${(1-\beta)}^2$ term is diminishingly small and the
deletion phase is quick.  For small values of $\beta$ (around
$5\%-10\%$) and typical values of $R \in [64, 128]$, $R\beta \gg 1$
and hence it dominates the value of the expression. Since $N
\beta=|D|$, the time complexity becomes directly proportional to the
size of the delete list.

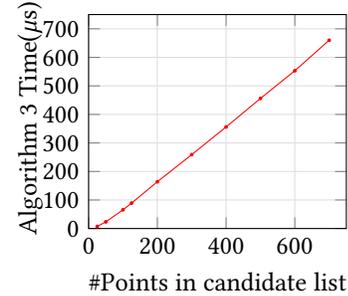
\begin{figure}[ht]
	\begin{center}
	\begin{tikzpicture}
	\begin{axis}[
			  scale=0.5,
			  grid=major, % Display a grid
			  grid style={solid,gray!25}, % Set the style
			  xmin = 0, xmax = 750,
			  ymin = 0, ymax = 750,
			  xlabel=\#Points in candidate list, % Set the labels
			  ylabel=~\Cref{alg:robustprune} Time($\mu$s),
			  ylabel shift=-7pt,                   
			  ytick distance = 100,
			]
			\addplot [red, mark = *, mark size = 0.5pt]table[x=X,y=Y,col sep=space] {PruneTimeSIFT1M.txt}; 
	\end{axis}
	\end{tikzpicture}
	\caption{Trend of ~\Cref{alg:robustprune} run time with increasing size of candidate list when 10\% of the SIFT1M index is being deleted.}
	\label{fig:prune-procedure-time}
	\end{center}
	\end{figure}

We demonstrate the linear time complexity of ~\Cref{alg:robustprune} in ~\Cref{fig:prune-procedure-time}. 
We delete a small fraction(10\%) of SIFT1M \algo index and record the time taken by ~\Cref{alg:robustprune} as the candidate list size increases.

\section{\recall{k}{k} for various k values} \label{sec:krecallk_appendix}
\subsection{\algofresh}
	\subsubsection{Search Latency vs Recall}
	In ~\Cref{fig:norm-query-latency-k-1,fig:norm-query-latency-k-10,fig:norm-query-latency-k-100}, we compare the search latencies for \algo and build time-normalized \algofresh (build parameters adjusted to match the build time of \algo) for various \recall{k}{k}.
	For \recall{1}{1}  and \recall{10}{10}, we compare latencies for 95\%, 98\% and 99\% recall. 
	For \recall{100}{100}, we compare over 98\% and 99\% recall because the lowest search list parameter $L$ value gives 98\% recall.
	\begin{figure}[t]
    \begin{center}
    \begin{tikzpicture}
    \pgfplotstableread{
    0.0866922	0.114615
    0.152367	0.187775
    0.266392	0.339833
    0.0664649	0.0950075
    0.10547	    0.138622
    0.153084	0.190855
    1.4888	    1.50512
    2.86701     2.57573
    4.11502	    3.6439
    1.61801     1.73028
    1.81611	    2.98287
    2.01109	    2.24742
    }\datatable
    
    \node [align=left, font=\small,
        text width=1.5cm, inner sep=0.25cm] at (1.18cm, -0.55cm) {\textsc{SIFT1M}};
    
    \node [align=left, font=\small,
    text width=1.5cm, inner sep=0.25cm] at (2.75cm, -0.55cm) {\textsc{DEEP1M}};
    
    \node [align=left, font=\small,
    text width=1.5cm, inner sep=0.25cm] at (4.32cm, -0.55cm) {\textsc{GIST1M}};
    
    \node [align=left, font=\small,
    text width=1.5cm, inner sep=0.25cm] at (5.9cm, -0.55cm) {\textsc{SIFT100M}};
    
    \begin{axis}[
       ybar,
        title style={align=center},
        ticks=both,
        ytick={1.00, 2.00, 3.00, 4.00, 5.0},
        axis x line = bottom,
        axis y line = left,
        axis line style={-|},
        nodes near coords = \rotatebox{90}{{\pgfmathprintnumber[fixed zerofill, precision=2]{\pgfplotspointmeta}}},
        nodes near coords align={vertical},
        every node near coord/.append style={font=\footnotesize, fill=none, yshift=0.5mm},
        enlarge y limits={lower, value=0.1},
        enlarge y limits={upper, value=0.22},
        ylabel=Query latency(ms),
       xtick=data,
       /pgf/bar width=4pt,
        ymin = 0,
        ymajorgrids,
        xticklabels={ 
            $95$, 
            $98$,
            $99$,
            $95$, 
            $98$,
            $99$,
            $95$, 
            $98$,
            $99$,
            $95$, 
            $98$,
            $99$},
        legend style={at={(0.5, -0.30)}, anchor=north, legend columns=3},
        every axis legend/.append style={nodes={right}, inner sep = 0.1cm},
       x tick label style={align=center, yshift=0cm, font=\footnotesize},
       y tick label style={xshift=0.0cm, font=\small},
        enlarge x limits=0.1,
        width=0.5\textwidth,
        height=5cm,
    ]
    \pgfplotsinvokeforeach {0,...,1}{
        \addplot table [x expr={\coordindex-mod(\coordindex, 3)/6}, y index=#1] {\datatable};
    }
    
    \draw (axis cs:2.33,0) -- ({axis cs:2.33,0}|-{rel axis cs:0.5,1});
    \draw (axis cs:5.33,0) -- ({axis cs:5.33,0}|-{rel axis cs:0.5,1});
    \draw (axis cs:8.33,0) -- ({axis cs:8.33,0}|-{rel axis cs:0.5,1});
    % \draw (axis cs:11.33,0) -- ({axis cs:11.33,0}|-{rel axis cs:0.5,1});
    \legend{\algo \hspace*{8pt}, \algofresh} 
    \end{axis}
    \end{tikzpicture}
    \end{center}
    \vspace{-6pt}
    \caption{Query latency for \algo and build-time normalized \algofresh \recall{1}{1} at 95\%, 98\%, and 99\%.
    }
    \label{fig:norm-query-latency-k-1}
    \vspace{-12pt}
    \end{figure}

    \begin{figure}[t]
        \begin{center}
        \begin{tikzpicture}
        \pgfplotstableread{
        0.105005	0.145307
        0.156845	0.221025
        0.212898	0.296373
        0.0923057	0.124106
        0.149137	0.191284
        0.208537	0.26076
        1.7799	    2.09867
        3.06158     3.59773
        4.71566	    5.28611
        1.76531     1.81545
        2.09505	    2.47668
        2.24816	    2.91166
        }\datatable
        
        \node [align=left, font=\small,
            text width=1.5cm, inner sep=0.25cm] at (1.18cm, -0.55cm) {\textsc{SIFT1M}};
        
        \node [align=left, font=\small,
        text width=1.5cm, inner sep=0.25cm] at (2.75cm, -0.55cm) {\textsc{DEEP1M}};
        
        \node [align=left, font=\small,
        text width=1.5cm, inner sep=0.25cm] at (4.32cm, -0.55cm) {\textsc{GIST1M}};
        
        \node [align=left, font=\small,
        text width=1.5cm, inner sep=0.25cm] at (5.9cm, -0.55cm) {\textsc{SIFT100M}};
        
        \begin{axis}[
           ybar,
            title style={align=center},
            ticks=both,
            ytick={1.00, 2.00, 3.00, 4.00, 5.00, 6.00},
            axis x line = bottom,
            axis y line = left,
            axis line style={-|},
            nodes near coords = \rotatebox{90}{{\pgfmathprintnumber[fixed zerofill, precision=2]{\pgfplotspointmeta}}},
            nodes near coords align={vertical},
            every node near coord/.append style={font=\footnotesize, fill=none, yshift=0.5mm},
            enlarge y limits={lower, value=0.1},
            enlarge y limits={upper, value=0.22},
            ylabel=Query latency(ms),
           xtick=data,
           /pgf/bar width=4pt,
            ymin = 0,
            ymajorgrids,
            xticklabels={ 
                $95$, 
                $98$,
                $99$,
                $95$, 
                $98$,
                $99$,
                $95$, 
                $98$,
                $99$,
                $95$, 
                $98$,
                $99$},
            legend style={at={(0.5, -0.30)}, anchor=north, legend columns=3},
            every axis legend/.append style={nodes={right}, inner sep = 0.1cm},
           x tick label style={align=center, yshift=0cm, font=\footnotesize},
           y tick label style={xshift=0.0cm, font=\small},
            enlarge x limits=0.1,
            width=0.5\textwidth,
            height=5cm,
        ]
        \pgfplotsinvokeforeach {0,...,1}{
            \addplot table [x expr={\coordindex-mod(\coordindex, 3)/6}, y index=#1] {\datatable};
        }
        
        \draw (axis cs:2.33,0) -- ({axis cs:2.33,0}|-{rel axis cs:0.5,1});
        \draw (axis cs:5.33,0) -- ({axis cs:5.33,0}|-{rel axis cs:0.5,1});
        \draw (axis cs:8.33,0) -- ({axis cs:8.33,0}|-{rel axis cs:0.5,1});
        % \draw (axis cs:11.33,0) -- ({axis cs:11.33,0}|-{rel axis cs:0.5,1});
        \legend{\algo \hspace*{8pt}, \algofresh} 
        \end{axis}
        \end{tikzpicture}
        \end{center}
        \vspace{-6pt}
        \caption{Query latency for \algo and build-time normalized \algofresh \recall{10}{10} at 95\%, 98\%, and 99\%.
        }
        \label{fig:norm-query-latency-k-10}
        \vspace{-12pt}
        \end{figure}

        \begin{figure}[t]
            \begin{center}
            \begin{tikzpicture}
            \pgfplotstableread{
            0.325122	0.492132
            0.422676	0.581924
            0.306212	0.423835
            0.422793	0.571766
            5.93945     5.6913
            7.56424	    7.84618
            3.11993	    3.3255
            3.7814	    4.2541
            }\datatable
            
            \node [align=left, font=\small,
                text width=1.5cm, inner sep=0.25cm] at (1.18cm, -0.55cm) {\textsc{SIFT1M}};
            
            \node [align=left, font=\small,
            text width=1.5cm, inner sep=0.25cm] at (2.75cm, -0.55cm) {\textsc{DEEP1M}};
            
            \node [align=left, font=\small,
            text width=1.5cm, inner sep=0.25cm] at (4.32cm, -0.55cm) {\textsc{GIST1M}};
            
            \node [align=left, font=\small,
            text width=1.5cm, inner sep=0.25cm] at (5.9cm, -0.55cm) {\textsc{SIFT100M}};
            
            \begin{axis}[
               ybar,
                title style={align=center},
                ticks=both,
                ytick={1.00, 2.00, 3.00, 4.00, 5.00, 6.00, 7.00, 8.00},
                axis x line = bottom,
                axis y line = left,
                axis line style={-|},
                nodes near coords = \rotatebox{90}{{\pgfmathprintnumber[fixed zerofill, precision=2]{\pgfplotspointmeta}}},
                nodes near coords align={vertical},
                every node near coord/.append style={font=\footnotesize, fill=none, yshift=0.5mm},
                enlarge y limits={lower, value=0.1},
                enlarge y limits={upper, value=0.22},
                ylabel=Query latency(ms),
               xtick=data,
               /pgf/bar width=4pt,
                ymin = 0,
                ymajorgrids,
                xticklabels={  
                    $98$,
                    $99$, 
                    $98$,
                    $99$, 
                    $98$,
                    $99$, 
                    $98$,
                    $99$},
                legend style={at={(0.5, -0.30)}, anchor=north, legend columns=3},
                every axis legend/.append style={nodes={right}, inner sep = 0.1cm},
               x tick label style={align=center, yshift=0cm, font=\footnotesize},
               y tick label style={xshift=0.0cm, font=\small},
                enlarge x limits=0.1,
                width=0.5\textwidth,
                height=5cm,
            ]
            \pgfplotsinvokeforeach {0,...,1}{
                \addplot table [x expr={\coordindex-mod(\coordindex, 3)/6}, y index=#1] {\datatable};
            }
            
            \draw (axis cs:2.33,0) -- ({axis cs:2.33,0}|-{rel axis cs:0.5,1});
            \draw (axis cs:5.33,0) -- ({axis cs:5.33,0}|-{rel axis cs:0.5,1});
            \draw (axis cs:8.33,0) -- ({axis cs:8.33,0}|-{rel axis cs:0.5,1});
            % \draw (axis cs:11.33,0) -- ({axis cs:11.33,0}|-{rel axis cs:0.5,1});
            \legend{\algo \hspace*{8pt}, \algofresh} 
            \end{axis}
            \end{tikzpicture}
            \end{center}
            \vspace{-6pt}
            \caption{Query latency for \algo and build-time normalized \algofresh \recall{100}{100} at 98\%, and 99\%.
            }
            \label{fig:norm-query-latency-k-100}
            \vspace{-12pt}
            \end{figure}    

	\subsubsection{Recall stability of \algofresh}
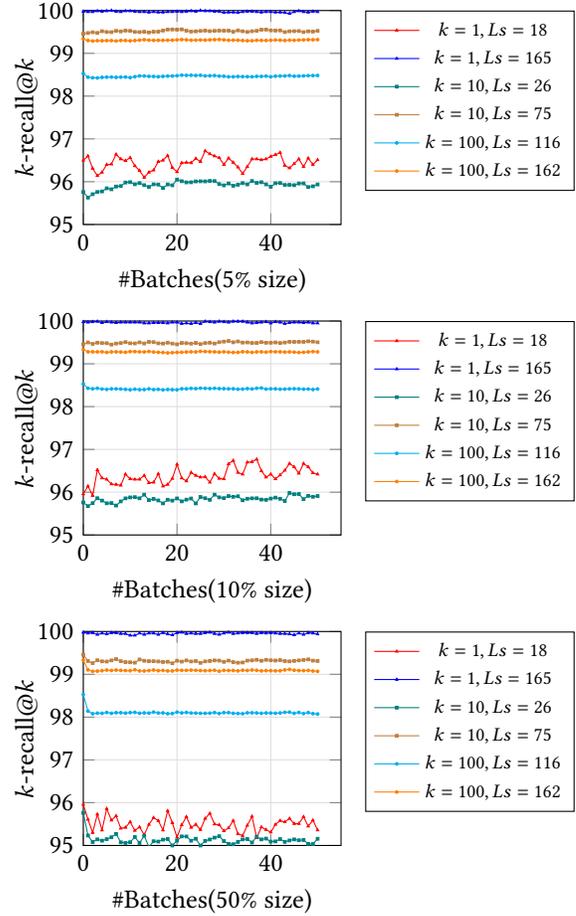
\begin{figure}[ht]
		\begin{center}
		\begin{tikzpicture}
		\begin{axis}[
				  scale=0.5,
				  grid=major, % Display a grid
				  grid style={solid,gray!25}, % Set the style
				  xmin = 0, xmax = 55,
				  ymin = 95, ymax = 100,
				  xlabel=\#Batches(5\% size), % Set the labels
				  ylabel=\recall{k}{k},
				  ylabel shift=-7pt,
				  ytick distance = 1,
				  legend columns = 1,
					legend style={at={(1.5,1)},anchor=north,font=\scriptsize},                   
				]
				\addplot [red, mark = triangle*, mark size = 0.5pt]table[x=X,y=Y1,col sep=space] {plots/RecallStability_k_del5_SIFT1M.txt}; 
				\addplot [blue, mark = triangle*, mark size = 0.5pt]table[x=X,y=Y2,col sep=space] {plots/RecallStability_k_del5_SIFT1M.txt}; 
				\addplot [teal, mark = square*, mark size = 0.5pt]table[x=X,y=Y3,col sep=space] {plots/RecallStability_k_del5_SIFT1M.txt};
				\addplot [brown, mark = square*, mark size = 0.5pt]table[x=X,y=Y4,col sep=space] {plots/RecallStability_k_del5_SIFT1M.txt};
				\addplot [cyan, mark = *, mark size = 0.5pt]table[x=X,y=Y5,col sep=space] {plots/RecallStability_k_del5_SIFT1M.txt};
				\addplot [orange, mark = *, mark size = 0.5pt]table[x=X,y=Y6,col sep=space] {plots/RecallStability_k_del5_SIFT1M.txt};
				\legend{$k=1,Ls= 18$\\ $k=1,Ls= 165$\\ $k=10,Ls= 26$\\ $k=10,Ls= 75$\\ $k=100,Ls= 116$\\ $k=100,Ls= 162$\\}
		\end{axis}
		\end{tikzpicture}
	
		\begin{tikzpicture}
		\begin{axis}[
				  scale=0.5,
				  grid=major, % Display a grid
				  grid style={solid,gray!25}, % Set the style
				  xmin = 0, xmax = 55,
				  ymin = 95, ymax = 100,
				  xlabel=\#Batches(10\% size), % Set the labels
				  ylabel=\recall{k}{k},
				  ylabel shift=-7pt,
				  ytick distance = 1,
				  legend columns = 1,
          		  legend style={at={(1.5,1)},anchor=north,font=\scriptsize},                   
				]
				\addplot [red, mark = triangle*, mark size = 0.5pt]table[x=X,y=Y1,col sep=space] {plots/RecallStability_k_del10_SIFT1M.txt}; 
				\addplot [blue, mark = triangle*, mark size = 0.5pt]table[x=X,y=Y2,col sep=space] {plots/RecallStability_k_del10_SIFT1M.txt}; 
				\addplot [teal, mark = square*, mark size = 0.5pt]table[x=X,y=Y3,col sep=space] {plots/RecallStability_k_del10_SIFT1M.txt};
				\addplot [brown, mark = square*, mark size = 0.5pt]table[x=X,y=Y4,col sep=space] {plots/RecallStability_k_del10_SIFT1M.txt};
				\addplot [cyan, mark = *, mark size = 0.5pt]table[x=X,y=Y5,col sep=space] {plots/RecallStability_k_del10_SIFT1M.txt};
				\addplot [orange, mark = *, mark size = 0.5pt]table[x=X,y=Y6,col sep=space] {plots/RecallStability_k_del10_SIFT1M.txt};
				\legend{$k=1,Ls=18$\\ $k=1,Ls=165$\\ $k=10,Ls= 26$\\ $k=10,Ls= 75$\\ $k=100,Ls= 116$\\ $k=100,Ls= 162$\\}
		\end{axis}
		\end{tikzpicture}
	
			\begin{tikzpicture}
			\begin{axis}[
					  scale=0.5,
					  grid=major, % Display a grid
					  grid style={solid,gray!25}, % Set the style
					  xmin = 0, xmax = 55,
					  ymin = 95, ymax = 100,
					  xlabel=\#Batches(50\% size), % Set the labels
					  ylabel=\recall{k}{k},
					  ylabel shift=-7pt,
					  ytick distance = 1,
					  legend columns = 1,
						legend style={at={(1.5,1)},anchor=north,font=\scriptsize},                   
					]
					\addplot [red, mark = triangle*, mark size = 0.5pt]table[x=X,y=Y1,col sep=space] {plots/RecallStability_k_del50_SIFT1M.txt}; 
					\addplot [blue, mark = triangle*, mark size = 0.5pt]table[x=X,y=Y2,col sep=space] {plots/RecallStability_k_del50_SIFT1M.txt}; 
					\addplot [teal, mark = square*, mark size = 0.5pt]table[x=X,y=Y3,col sep=space] {plots/RecallStability_k_del50_SIFT1M.txt};
					\addplot [brown, mark = square*, mark size = 0.5pt]table[x=X,y=Y4,col sep=space] {plots/RecallStability_k_del50_SIFT1M.txt};
					\addplot [cyan, mark = *, mark size = 0.5pt]table[x=X,y=Y5,col sep=space] {plots/RecallStability_k_del50_SIFT1M.txt};
					\addplot [orange, mark = *, mark size = 0.5pt]table[x=X,y=Y6,col sep=space] {plots/RecallStability_k_del50_SIFT1M.txt};
					\legend{$k=1,Ls= 18$\\ $k=1,Ls= 165$\\ $k=10,Ls= 26$\\ $k=10,Ls= 75$\\ $k=100,Ls= 116$\\ $k=100,Ls= 162$\\}
			\end{axis}
			\end{tikzpicture}
			\caption{Post-insertion search \recall{k}{k} for $k=1,10,100$ of \algofresh index over 50 cycles of deletion
			and re-insertion of 5\%, 10\% and 50\% (rows 1, 2 and 3 respectively) of SIFT1M index with
			varying search list size parameter $L$.}
			\label{fig:recall-stability-k}
			\end{center}
			\end{figure}

	In ~\Cref{fig:recall-stability-k}, we demonstrate \recall{k}{k} stability of \algofresh for commonly used k values. 
	We show the post-insertion recall trends for \recall{1}{1}, \recall{10}{10} and \recall{100}{100}. 
	For $k = 1$, we show how the 95\% and 99.9\% recall are stable. 
	For $k = 10$, we show that 95\% and 99\% recall are stable.
	For $k = 100$, the lowest valid search list parameter $L$ value is 100 and this gives 98\% recall. 
	So we show the stability of 98\% and 99\% recall.

	\begin{figure}[t]
		\begin{center}
			\begin{tikzpicture}
				\begin{axis}[
					scale=0.75,
					grid=major, % Display a grid
					grid style={solid,gray!25}, % Set the style
					xmin = 0, xmax = 12000,
					ymin = 20, ymax = 70,
					xlabel= Time elapsed(sec), % Set the labels
					ylabel=Search latency(ms),
					ylabel shift=-7pt,
					xlabel shift =20pt,
					ytick distance = 10
					]
					\addplot [red, mark = none, mark size = 0.5pt]table[x=X,y=Y,col sep=space] {appendix_plots/SS_1b_recall100_latency.txt}; 
					\filldraw[fill=green!40!white, fill opacity=0.2, draw=none] (0,0) rectangle (8929,70);
					\filldraw[fill=blue!40!white, fill opacity=0.2, draw=none] (8929,0) rectangle (10147,70);
					\filldraw[fill=brown!40!white, fill opacity=0.2, draw=none] (10147,0) rectangle (10687,70);
					\node[align=center, fill=white] at (4400, 50) {Delete};
					\node[align=center, fill=white] at (9000, 50) {Insert};
					\node[align=center, fill=white] at (10400, 60) {Patch};
					\draw[dashed] (8929, 0) -- (8929, 70);
					\draw[dashed] (10147, 0) -- (10147, 70);
					\draw[dashed] (10687, 0) -- (10687, 70);
				\end{axis}
			\end{tikzpicture}
			\caption{Trend of mean search latencies for 95\% search
				\recall{100}{100}, zoomed in over one cycle of inserting and deleting 30M points
				concurrently into a 800M SIFT index, using different 10 for search. Each point is the mean latency over a search batch of 10000 queries.}
			\label{fig:steady-800m-latency-recall-100}
		\end{center}
	\end{figure}

	\begin{figure}[t]
		\begin{center}
			\begin{tikzpicture}
				\begin{axis}[
					scale=0.75,
					grid=major, % Display a grid
					grid style={solid,gray!25}, % Set the style
					xmin = 0, xmax = 12000,
					ymin = 5, ymax = 30,
					xlabel= Time elapsed(sec), % Set the labels
					ylabel=Search latency(ms),
					ylabel shift=-7pt,
					xlabel shift =20pt,
					ytick distance = 5
					]
					\addplot [red, mark = none, mark size = 0.5pt]table[x=X,y=Y,col sep=space] {appendix_plots/SS_1b_recall10_latency.txt}; 
					\filldraw[fill=green!40!white, fill opacity=0.2, draw=none] (0,0) rectangle (8929,30);
					\filldraw[fill=blue!40!white, fill opacity=0.2, draw=none] (8929,0) rectangle (10147,30);
					\filldraw[fill=brown!40!white, fill opacity=0.2, draw=none] (10147,0) rectangle (10687,30);
					\node[align=center, fill=white] at (4400, 20) {Delete};
					\node[align=center, fill=white] at (9000, 20) {Insert};
					\node[align=center, fill=white] at (10400, 25) {Patch};
					\draw[dashed] (8929, 0) -- (8929, 30);
					\draw[dashed] (10147, 0) -- (10147, 30);
					\draw[dashed] (10687, 0) -- (10687, 30);
				\end{axis}
			\end{tikzpicture}
			\caption{Trend of mean search latencies for 95\% search
				\recall{10}{10}, zoomed in over one cycle of inserting and deleting 30M points
				concurrently into a 800M SIFT index, using different 10 for search. Each point is the mean latency over a search batch of 10000 queries.}
			\label{fig:steady-800m-latency-recall-10}
		\end{center}
	\end{figure}

	\subsection{\diskanntwo}
	\subsubsection{Search latencies over one merge cycle}
	In ~\Cref{fig:steady-800m-latency-recall-100,fig:steady-800m-latency-recall-10}, we present the 
	evolution of mean search latency for \recall{100}{100} and \recall{10}{10} over 
	the course of one merge cycle in a 800M \diskanntwo steady-state experiment.
	
\begin{figure}[t]
		\begin{center}
			\begin{tikzpicture}
				\begin{axis}[
					scale=0.5,
					grid=major, % Display a grid
					grid style={solid,gray!25}, % Set the style
					xmin = 0, xmax = 35,
					ymin = 2, ymax = 10,
					xlabel= Number of threads, % Set the labels
					ylabel=Search latency(ms),
					ylabel shift=-7pt,
					xlabel shift =10pt,
					]
					\addplot [red, mark = *, mark size = 0.5pt]table[x=X,y=Y,col sep=space] {appendix_plots/MeanSearchLatencyz840_800M.txt}; 
					%\addplot [blue, mark = *, mark size = 0.5pt]table[x=X,y=Y,col sep=space] {appendix_plots/SearchLatency99z840_800M.txt}; 
					%\legend{Mean Latency, $99^{th}$ percentile latency}
				\end{axis}
			\end{tikzpicture}
			\caption{Trend of mean latencies for $95\%$ search
				recall on a 800M SIFT index with different number of threads. Each point is calculated over a search batch of 10000 queries}
			\label{fig:800m-search-latency}
		\end{center}
	\end{figure}
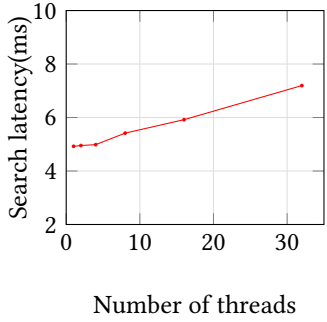	

\section{Search latency of \diskanntwo}
In ~\Cref{fig:800m-search-latency}, we observe the effect of number of search threads on mean search latencies for 800M index when no merge is going on.

\begin{figure}[t]
	\begin{center}
		\begin{tikzpicture}
			\begin{axis}[
				scale=0.75,
				grid=major, % Display a grid
				grid style={solid,gray!25}, % Set the style
				xmin = 0, xmax = 16000,
				ymin = 5, ymax = 30,
				xlabel= Time elapsed(sec), % Set the labels
				ylabel=Search latency(ms),
				ylabel shift=-7pt,
				xlabel shift =20pt,
				ytick distance = 5,
				legend style={at={(1.2, 0.8)},anchor=north},
				]
				\addplot [red, mark = none, mark size = 0.5pt]table[x=X,y=Y1,col sep=space] {appendix_plots/SS_800m_40_merge_1_search_latency.txt}; 
				\addplot [blue, mark = none, mark size = 0.5pt]table[x=X,y=Y1,col sep=space] {appendix_plots/SS_800m_40_merge_2_search_latency.txt}; 
				\addplot [green, mark = none, mark size = 0.5pt]table[x=X,y=Y1,col sep=space] {appendix_plots/SS_800m_40_merge_4_search_latency.txt}; 
				\addplot [orange, mark = none, mark size = 0.5pt]table[x=X,y=Y1,col sep=space] {appendix_plots/SS_800m_40_merge_6_search_latency.txt}; 
				\filldraw[fill=green!40!white, fill opacity=0.2, draw=none] (0,0) rectangle (7751,25);
				\filldraw[fill=blue!40!white, fill opacity=0.2, draw=none] (7751,0) rectangle (12016,25);
				\filldraw[fill=brown!40!white, fill opacity=0.2, draw=none] (12016,0) rectangle (13314,25);
				\node[align=center, fill=white] at (3500, 20) {Delete};
				\node[align=center, fill=white] at (9900, 20) {Insert};
				\node[align=center, fill=white] at (12500, 28) {Patch};
				\draw[dashed] (7751, 0) -- (7751, 25);
				\draw[dashed] (12016, 0) -- (12016, 25);
				\draw[dashed] (13314, 0) -- (13314, 25);
				\legend{1,2,4,6}
			\end{axis}
		\end{tikzpicture}
		\caption{Trend of mean search latencies for 92\% search
			recall, zoomed in over one cycle of inserting and deleting 30M points
			concurrently into a 800M SIFT index, using different number of threads for search. Each point is the mean latency over a search batch of 10000 queries.}
		\label{fig:steady-800m-fixed-merge-threads}
	\end{center}
\end{figure}
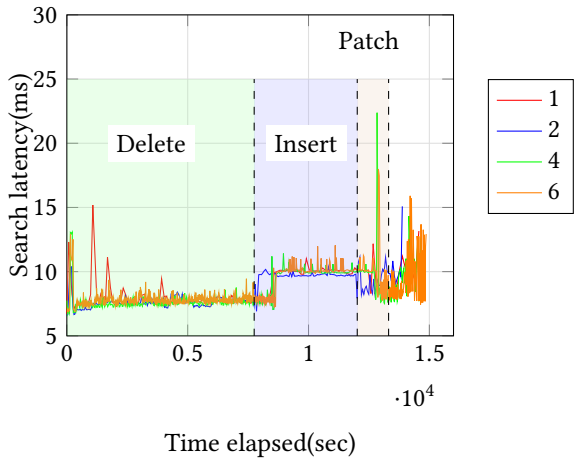

\section{Concurrency during \merger}
	In this section, we present our observations on search latency during merge through in-depth experiments on \diskanntwo merge with varying thread allocations. 
	All experiments are 30M insertions and deletions into a 800M \diskanntwo index.
		
	\subsection{Search threads fixed - varying merge threads}
	We run the merge on SIFT800M index with different thread allocations to understand the effect of merge on search latency. 
	In ~\Cref{fig:steady-800m-fixed-search-threads}, we plot a smoothed curve of mean search latencies when merge uses 20 and 40 threads. 
	Merge with 40 threads takes approximately half the time as that with 20, so there are two x axes adjusted to roughly align their Delete, Insert and Patch phases.
	As evident from the figure, search latencies with 40 thread merge are consistently higher in the Delete and Insert phases of merge.

	\subsection{Merge threads fixed - varying search threads}
	We run the merge on SIFT800M index with different thread allocations to understand the effect of number of search threads used during merge on search latency. 
	We increase the number of search threads while fixing 40 threads for merge, and observe how the search latency trend evolves in over one merge cycle ~\Cref{fig:steady-800m-fixed-merge-threads}.

\end{document}